\documentclass[useAMS,usenatbib]{mn2e}
\usepackage{graphicx,color,epsfig}
\usepackage{color,epsfig}

\newcommand{\be}{\begin{equation}}
\newcommand{\ee}{\end{equation}}

\newcommand{\apj}{ApJ}

\newcommand{\mnras}{MNRAS}
\newcommand{\aap}{A\&A}
\newcommand{\araa}{ARA\&A}
\newcommand{\apjl}{ApJL}

\newcommand{\nat}{Nature}

\def\ltsima{$\; \buildrel < \over \sim \;$}
\def\simlt{\lower.5ex\hbox{\ltsima}}
\def\gtsima{$\; \buildrel > \over \sim \;$}
\def\simgt{\lower.5ex\hbox{\gtsima}}

\newcommand\rfc{R_{\rm fc}}
\newcommand\mfc{M_{\rm fc}}

\def\msun{{\,{\rm M}_\odot}}

\newcommand\mearth{{\,{\rm M}_{\oplus}}}

\def\del#1{{}}

\title[Solid  cores inside giant planet embryos]{Formation
  of terrestrial planet cores inside giant planet embryos.}

\author[S. Nayakshin]{Sergei Nayakshin \\ Department of Physics \& Astronomy,
  University of Leicester, Leicester, LE1 7RH, UK}

\begin{document}

\date{Accepted 2008 ?? ??. Received 2008 ?? ??; in original form 2008 05 ??}

\pagerange{\pageref{firstpage}--\pageref{lastpage}} \pubyear{2008}

\maketitle

\label{firstpage}

\begin{abstract}
  Giant planet embryos are believed to be spawned by gravitational
  instability in massive extended ($R\sim 100$ AU) protostellar
  discs. In a recent paper we have shown that dust can sediment inside
  the embryos, as argued earlier by \cite{Boss98} in a slightly
  different model. Here we study numerically the next stage of this
  process -- the formation of a solid core. If conditions are
  conducive to solid core formation, the centre of the gas cloud goes
  through the following sequence of phases: (i) becomes grain (and
  metal) rich; (ii) forms a terrestrial mass solid core via a rapid
  collapse driven by self-gravity of the grains; (iii) starts to
  accrete a gaseous atmosphere when the solid core reaches mass of a
  few to $10 \mearth$. This sequence of events may build either
  terrestrial planet cores or metal rich giant planets inside the
  larger gas reservoir of the giant planet embryo. In a companion
  Letter we argue that tidal and irradiation effects from the parent
  star should disrupt the outer metal poor layers of the embryo,
  releasing nearly ``ready to use'' planets.  We propose this as an
  alternative way to build planets.
\end{abstract}

\begin{keywords}
{}
\end{keywords}

\section{Introduction}\label{intro}

It would be clearly an attractive proposition if planets could form in
a way similar to stars, e.g., by gravitational contraction followed by
a rapid collapse of gas or solid material clumps. A clear route to
this is gravitational instability (GI) of the parent protoplanetary
disc, when the latter fragments onto gaseous clumps with masses of a
few Jupiter mass ($M_J = 10^{-3} \msun$).  \cite{Boss97} argued that
giant planets may form this way. However giant planets in the Solar
System contain massive solid cores, and they are also much more metal
rich than the Sun. Thus a simple gravitational collapse of a giant
planet embryo could not work. \cite{Boss98} suggested that solid cores
of the order of a few Earth masses ($\mearth \approx 3\times 10^{-6}
\msun$) can be formed if dust grows and sediments inside the giant
embryos. \cite{BossEtal02} further suggested that even more massive
and icy cores can be built inside giant proto-planets at 10 to a few
tens of AU distances. These authors further pointed out that
metalicity of such planets could be raised if gas envelopes of giant
planet embryos are removed by photo-evaporation due to a nearby OB
star. Such a model would account for both the solid cores and high
metalicities of the giant planets.

However, \cite{WuchterlEtal00} argued that giant planet embryos may
become convective within the first $\sim 100$ years of their
evolution, and that convection may inhibit grain
sedimentation. \cite{HelledEtal08,HS08} have recently confirmed this,
obtaining much smaller solid cores than the model of \cite{Boss98}
predicts. These authors also found that their model gas clumps either
were too hot to begin with or became hot too soon for a significant
dust growth, instead vaporising the grains. 

Another objection to the gravitational instability model for planet
formation was pointed out by \cite{Rafikov05} who showed analytically
that the disc cannot even fragment on self-bound gas clumps inside
$\sim 30$ AU at least.  The core accretion (CA) model for planet
formation \citep[e.g.,][]{Wetherill90,PollackEtal96,IdaLin08}
therefore seems more attractive to many.

However, the extrasolar planets observed at large ($R_p \sim 100$ AU)
separations from their parent stars \citep{BaraffeEtal10} cannot be
explained by the CA model due to exceedingly long core formation
times. These planets were almost certainly formed by the gravitational
disc instability \citep[e.g., see simulations
by][]{SW08,InutsukaEtal09,MachidaEtal10,RiceEtal10}.

Formation by GI is not absolutely forbidden even for the giant planets
observed inside a few AU disc. True, the planets could not have started
to form in situ there because initially giant embryos are much less
dense than the local tidal density (see \S 2 below).  However, radial
migration of massive planets \citep{GoldreichTremaine80} due to
planet-disc torques is now an accepted ingredient of any planet
formation theory, and provides an attractive explanation for the
observed distribution of extrasolar planets in the radial range of
$\sim 0.1-2.5 $ AU \citep{Armitage07}, and ``hot jupiters'' even
closer in. We do not see why it is not possible for the giant planets
to form at very large radii by GI and then migrate inwards to a few AU
if they contract quickly enough. In a companion paper (Nayakshin
2010b; submitted to MNRAS, referred below as paper III) we argue that
such a long-range migration is possible in the very early phase of
formation of the parent star, when the disc mass is comparable to that
of the star.

Furthermore, Boss et al, Wuchterl et al, and Helled and co-authors all
considered embryos with gas densities of order $10^{-8}$ g cm$^{-3}$
(perhaps having in mind the tidal density constraints for Jupiter at
$\sim$ 5 AU). Nayakshin (2010a, ``paper I'') showed the disc
fragmentation constrains \citep{Gammie01,Rice05,Rafikov05} require
that the initial density of gas clumps formed at $\sim 100$ AU from
the star is around $10^{-13}$ g cm$^{-3}$. This fluffy initial state
of giant embryos is also much cooler, with typical temperatures of
only $\sim 100$ K, an order of magnitude lower than in the study by
Helled an co-authors.

It is thus not surprising that we found in paper I such embryos to be
much more promising sites of grain growth and sedimentation than did
\cite{HelledEtal08,HS08}. For reasonable opacity values, dust grows
and sediments inside isolated giant planet embryos with masses between
a few and $10-20$ Jupiter masses most readily. Since the embryos are
approximately isentropic at birth, convection did not appear to be
important in opposing grain sedimentation.

For technical reasons, the calculations were stopped in paper I when
the grain density in the inner region of the gas clump reached that of
the gas.  In this paper we extend the calculations similar to those
presented in paper I to later times. As predicted analytically in
paper I, the inner region becomes completely grain-dominated by mass
and then undergoes a rapid collapse. The gas component is
``protected'' against the collapse by the pressure gradient, and thus
the two phases separate at the cloud centre. The grains form a high
density ``solid''' core (it is most likely melted or even liquid given
the high energy release, but we shall still refer to it as a solid
core as it may solidify later on when it cools).

Our main results can be summarised as following:

\begin{itemize}

\item The heat generated by the rapid assembly of the high-Z core
  produces a negative feedback on the core's growth. Grains can be
  melted, driven away convectively or by a bulk gas expansion, cutting
  off further core growth. Thus, concerning the final mass of the
  solid cores, our results are less optimistic than that of
  \cite{Boss98,BossEtal02}, but more optimistic than that of
  \cite{HelledEtal08,HS08}. Continuing the trends from paper I, the
  embryos of a few to ten Jupiter masses usually yield the heaviest
  solid cores. More massive giant embryos are hotter to begin with and
  hence reach the grain vaporisation temperature sooner.

\item Gas immediately adjacent to the solid core is strongly
  influenced by the energy release due to the core's assembly. Strong
  convection sets in (thus confirming its importance but only after
  the core formation), frequently supplemented by an outward gas
  expansion.

\item Resolving the region near the solid core is challenging
  numerically, which implies that we probably underestimate the degree
  to which the gas becomes bound to the core. Even so, in a simulation
  where most of the giant embryo gets unbound, the solid core is left
  with a gas atmosphere of $\sim 0.03 \mearth$. This shows that solid
  cores bred inside giant embryos may attract a gaseous atmosphere
  from within the embryo. This atmosphere is metal rich.

\item We then use analytical arguments and a physical
  analogy to the classical results from the core accretion model for giant
  planet formation \citep[e.g.,][]{Mizuno80,PollackEtal96}. We suggest that as
  the solid core mass increases above $\sim 10-20 \mearth$, a runaway gas accretion
  sets in, in which the gas envelope collapses on itself. The resulting solid
  core plus a metal-rich self-bound atmosphere should be similar to a giant
  planet.

\item In one of the simulations presented below most of the gas
  envelope gets unbound by the energy release from the core. This
  shows that, at least in the high optical depth cases, the heat
  released by the core's accretion may remove the envelope without an
  outside help. We believe that this outer layers blowout process may
  be even more prominent (e.g., likely) when the massive atmosphere
  collapse occurs, as that may release even more energy.

\item We suggest that the outer metal-poor layers of the giant embryo  could
  also be removed by physical collisions of embryos.

\item We find that the protective giant embryo environment strongly
  diminishes the role of turbulence. We show that turbulent gas
  velocities would have to be as large as 0.3 times the sound speed to
  prevent gravitational collapse of the ``grain sphere'' into the
  solid core. As these high speeds are unlikely, we argue that solid
  cores can be built by a direct gravitational collapse inside the
  giant envelopes {\em without} having to go through the intermediate
  step of planetesimals. The planetesimals are thus not building
  blocks of planets in this picture. They are the material that did
  not join the core.

\item Thus, in contrast to the classical planetesimal core assembly
  route \citep[e.g.,][]{Safronov69,Weiden80,Wetherill90}, we would
  expect a far smaller fraction of the solid mass to be locked into
  intermediate (e.g., a km) size bodies. 

\item The main building blocks of solid cores in our scenario are
  pebble sized grains (e.g., $\sim 10$ cm). Grains of this size
  sediment efficiently enough and yet slowly enough as to avoid
  shattering in high speed collisions.

\item Concluding, our key point is that the combination of dust sedimentation,
  solid core accretion and gas envelope collapse, on the one hand, with the
  removal of the outer metal-poor envelope of the embryo on the other, may
  result in terrestrial planet embryos, terrestrial planets, or giant planets,
  depending on when the envelope removal occurs.

\end{itemize}

In the follow-up paper already mentioned above (paper III), we argue that
removal of the giant embryo by tidal shear from the star helped by heating
caused by irradiation is un-escapable for giant planets migrating closer in
than a few to a few ten AU. An exception to this could be a more massive giant
planet $M_p \simgt 20 M_J$ that could undergo the ``second collapse''
\citep{Larson69,Masunaga98} before it is disrupted. Such massive planets
should have metalicities close to that of their parent stars.

A hybrid regime of planet formation is possible, in which the solid
cores are hatched inside the giant planet embryos, delivered into the
inner disc and left there when the embryos are disrupted.  These solid
cores may continue their growth by accumulation of solid material from
the disc (rather than the parent embryo) in direct impacts, and then
accrete massive gaseous envelopes from the surrounding gas disc,
resulting in gas giant planets. In this case the beginning phase of
planet formation is as in the modified GI hypothesis we study, but
continuation as in the core accretion theory.

Future detailed calculations and comparison to observations are needed
to test feasibility of these ideas. In any event it seems very clear
that the gravitational instability model for planet formation has been
discounted by most authors without a proper critical review of all the
evidence.

\section{Preliminaries}\label{sec:1st}

\subsection{Initial properties of self-gravitating clumps}

In paper I we argued that the properties of the giant planet embryos formed by gravitational instability in the disc at their inception should be similar to that of the first cores of same  mass.  
The key properties of  first cores, taken from the results of \cite{MI99} as explained in
paper I, are briefly reproduced here for reader's convenience.
The opacity law is
taken in the form
\begin{equation}
\kappa(T) = \kappa_0 \left(\frac{T}{10}\right)^\alpha\;,
\label{kappa0}
\end{equation}
where $\kappa_0 $ is the opacity at $T= 10$ K, and reasonable values of
$\alpha$ are thought to be between 1 and 2. For simplicity, in this paper we
only consider the $\alpha=1$ case (\S \ref{sec:fiducial}). The initial radius of the first
core is given by 
\begin{equation}
\rfc = 6.0 \;\hbox{AU}\; m_1^{-1/3} T_1^{\frac{1+4\alpha}{9}}\kappa_{*}^{4/9}\;,
\label{rfc2}
\end{equation}
where $\kappa_{*} = \kappa_0/0.01$ and $T_1 = T_{\rm init}/10$K is the initial
gas temperature, e.g., the temperature of the parent gas material (the disc or
the molecular cloud). Finally, $m_1 = \mfc/10M_J$.  The mean density of the
first core is defined by
\begin{equation}
\rho_{\rm fc} = \frac{3 \mfc}{4\pi \rfc^3} = 6.6\times 10^{-12}\; \hbox{g
  cm}^{-3}\; m_1^{2} T_1^{-\frac{1+4\alpha}{3}}\kappa_{*}^{-4/3}\;,
\label{rho_mean}
\end{equation}
The mean temperature of the first core is estimated as
\begin{equation}
T = \frac{1}{3} \frac{G \mfc \mu}{k_B \rfc} = 146 \; \hbox{K}\;
m_1^{4/3} T_1^{-\frac{1+4\alpha}{9}}\kappa_{*}^{-4/9}\;.
\label{tvir_core}
\end{equation}
The cooling time of the first core is found to be 
\begin{equation}
t_{\rm cool} = 380 \; \hbox{years}\; m_1^{2/3} T_1^{-4/3} \kappa_{*}^{1/9}\;.
\label{tcool1}
\end{equation}
Note that the cooling time becomes longer with time, as the gas clump
contracts and becomes optically thicker (paper I).

\subsection{Dust sedimentation in isolated first cores}

The calculations in paper I were done under the assumption that the first core
gains or looses no mass from the surroundings. This limit applies if the mass
exchange rate satisfies (for $\alpha=1$) $|d\mfc/dt| \ll 10^{-2} M_J$
yr$^{-1}$, which does depend on parameters of the first core (see paper I). We
shall continue to assume this in the present paper as well, and hence the mass of the 
embryo is a free parameter of our calculations. 

It was shown in paper I that grain sedimentation timescales can be shorter
than the ``vaporisation time'', $t_{\rm vap}$, which is the time needed for
the first core to heat up to the grain vaporisation temperature of about 1400
K. The calculations were stopped when the density of the dust was equal to
that of the gas in the inner part of the cloud. Continuation of the
calculations leads to the dust density increasing even further, and then a
gravitational collapse. This then requires a treatment for the solid core
forming in the innermost part of the gas cloud. Before we present
calculations that now include formation of the cores, we first
discuss the order of magnitude of physical characteristics of the solid
cores.

\subsection{Maximum masses of solid cores}\label{sec:masses}

The first question to ask about the solid cores built inside isolated first
cores is how massive they could be.  The total mass of ``high-Z'' elements,
e.g., those excluding Hydrogen and Helium, is about
\begin{equation}
M_{\rm Z,t} = f_g \mfc \approx 60 \mearth\; \frac{f_g}{0.02} \; m_1\;,
\label{mzt}
\end{equation}
where $f_g = 0.02$ is the metal mass fraction (0.02 for gas of Solar
metalicity). This mass includes ices and organics that are probably too
volatile to contribute to the solid core, as the surrounding gas temperature
is typically around 1000 K. The mass of refractory material (e.g., silicates)
that can be used to build a massive core is hence expected to be about $\sim
1/3$ of the total heavy elements mass. Therefore, the maximum mass of the
solid core that can be built inside an {\em isolated} first core is
\begin{equation}
M_{\rm Z,c} = \frac{1}{3} f_g \mfc \approx 20 \mearth\; \frac{f_g}{0.02} \; m_1\;,
\label{mzc_max}
\end{equation}
The minimum core mass \citep[same as the opacity fragmentation
  limit][]{Low76,Rees76} is a few Jupiter masses, and the maximum mass that
permits grain sedimentation is typically a few tens of $M_J$. Accordingly, the total
mass of useful core-building material is expected to be in the range of $\sim 7$ to $\sim 60$
or so Earth's masses. These estimates should be tripled if the embryo is cool enough to permit 
formation of ices.

\subsection{Binding energy of solid cores}\label{sec:energy}

We expect the core of high-Z elements to have a density of the order of the
terrestrial planet densities, e.g., $\rho_c \sim $ a few g cm$^{-3}$. The
corresponding radial size of the solid core, $R_{\rm core} = (3M_{\rm c}/4\pi\rho_c)^{1/3}$,
is in the range of $10^8$ to a few times $10^9$ cm, which is about 5 orders of
magnitude smaller than the size of the first core. The binding energy of the
solid core is
\begin{equation}
E_{\rm bind, c} = \frac{3}{5} \frac{G M_{\rm core}^2}{R_{\rm core}} = 1.3 \times 10^{39} \;
\hbox{erg}\; \left(\frac{M_{\rm c}}{\mearth}\right)^{5/3} \rho_c^{-1/3} \;.
\label{ebind_p}
\end{equation}
The binding energy of the first core at formation is in the rough range of
$10^{40}$ to $10^{42}$ erg (see paper I), and it increases in proportion to
the virial temperature of the core (dimensionless $\tilde T(t)$ introduced in
paper I) as the core contracts. Hence a massive enough solid core, $M_p \sim 10
\mearth$, may release an amount of energy comparable to the binding energy of
the core itself, strongly affecting the core. Furthermore, for a constant
density gas sphere, inside the embryo, the binding energy scales as $M^2(R)/R \propto R^5$, and
the thermal energy scales as $M(R)\propto R^3$. Thus building up a massive
solid core is very likely to heat up the gas at least in the inner parts of
the first core.

\section{Numerical method}\label{sec:numerical}

\subsection{Overall approach}\label{sec:general}

We use the same set up, equations and numerical techniques as in paper I,
adding to the code several new features to describe the evolution of the solid
core and its effects on the surrounding gas.  In brief, spherically symmetric
Lagrangian coordinates are used. A cell is defined by its fixed gas mass,
whereas the radial position of the shell changes as the gas contracts or
expands. The equations are those widely used in stellar
evolution calculations, except we employ a time-dependent approach that does
not assume the hydrostatic or the energy balances for the gas. We find that
this time-dependent approach is very important for the problem at hand. Both
the grain and the gas distributions may evolve very rapidly when a heavy-Z
core is formed via a rapid (and hence quite luminous) accretion of grains.

The grains are described as a second fluid that is under the influence of
gravity and gas drag. Grains are allowed to slip through the gas, e.g., from
one gas radial shell into the neighbouring one. For example, grains can move
from an outer into an inner radial shell when they sediment, or be carried in
the opposite direction by turbulent or convective mixing.

\subsection{Convection and grain convective mixing}\label{sec:convection}

The transport of energy is primarily due to the classical radiation diffusion
flux, although convective flux does become important during certain stages of
the evolution, e.g., when the liquid core's luminosity is high. The mixing
length theory for convection is used \citep[e.g., \S 7 in][]{KW90}.
We impose an upper limit to the convective flux, $F_{\rm con}$,
given by $F_{\rm c,max} = c_s P$, where $c_s$ and $P$ are the gas sound speed
and pressure, respectively.

In contrast to paper I, the convective motions of the gas are allowed to
contribute to mixing of grains, impeding grain sedimentation. Treating this
process as a diffusion process, similar to the turbulent mixing \citep[paper I
  and][]{FromangPap06}, we ``add'' the convective mixing to turbulent mixing
by increasing the turbulent mixing coefficient $\alpha_d$ introduced in paper
I:
\begin{equation}
\alpha_d' = \alpha_d + \frac{F_{\rm con}}{F_{\rm c,max}}\;.
\end{equation}
This prescription is motivated by the fact that the convective mixing length
by definition becomes about equal to the gas pressure scale-height, e.g., the
radius of the first core, $\rfc$, when $F_{\rm con} = F_{\rm c,max}$. Note
that the maximum of $\alpha_d' \approx 1$. Strong convective motions could therefore be
quite important in opposing and perhaps even reversing grain sedimentation, but we
find that this fortuitously  occurs ``too late'', when a massive solid core has
already formed (see \S \ref{sec:10mj} below for an example).

\subsection{The inner grain boundary: a heavy elements
  core}\label{sec:core_formation}

Like previous authors \citep[e.g.,][]{HelledEtal08}, we found it very
challenging numerically to resolve the radial scales of the heavy-Z core for
the entire duration of the simulations. As there is very little gas mass on
these scales, except if and when a massive gas atmosphere builds up, many more
radial shells are needed. These shells require tiny time steps despite being
not very interesting for most of the calculation. Therefore, due to numerical
constrains, the length scales that we do resolve for most of the simulations
are still large compared with the solid core, of the order of $\sim 10^{11}$
cm. Note that a snapshot re-simulation can of course be done at a higher
resolution if one is interested in the small scale structure.

Given that the smallest scales are not properly resolved, there is an
uncertainty in choosing the boundary conditions for grains at $R=0$ radius,
that is in the very first {\em gas} mass zone. In paper I we assumed that the
grains in the innermost zone are suspended by turbulent mixing {\em until} the
grain density in the zone exceeded that of the gas. Presumably the inverse drag
effects imposed by grains on the gas stifle the turbulence once the
grains dominate the gas by mass. We stopped the simulations at that moment in
paper I, delaying a study of core accretion to this paper.

There is no a priory reason why turbulent mixing must be effective at
$R\rightarrow 0$. If it vanishes in that region, then grains may start to
sediment into a high-Z core even before they overwhelm the gas by mass in the
first zone.  \cite{HelledEtal08} classified a similar approach as ``case 1'',
whereas not introducing a solid core until the grains dominate the central zone is 
is analogous to their case 2. However, numerical experiments that we did showed  
that the difference in results between
these two cases is minor for most of the interesting parameter space. Namely,
we found that before the first zone is dominated by grains, the high-Z core
grows very slowly in ``case 1'' anyway. This is due to two factors. First of
all, the grain density in the central zone is small during these ``early''
times. Secondly, the grains are themselves small, yielding a rather small
sedimentation rate.

In the simulations we explored, there is a period of a very rapid (but
limited) growth of the solid core in both case 1 and 2. The only significant difference arises
for rapidly contracting first cores that reach grain vaporisation temperatures
before the rapid high-Z core growth phase. In case 1, a small high-Z core may be
built, whereas it is completely absent in case 2. We shall keep this fact
in mind, and proceed to use only the case 1 in this paper, since it turned out
to be numerically more stable during the episodes of the rapid high-Z core
growth. This does not lead to any loss of generality of our results. We still
cover the appropriate parameter space of the problem while avoiding
presentation of minor differences from very similar runs.

Therefore, our prescription for the core accretion rate is
\begin{equation}
\frac{d M_{\rm core}}{dt} = \cases{M_{d, 1} (u_{2}-u_{a2})/R_{g2} \qquad
    \hbox{if }\;  {u_{2}-u_{a2}} > 0 \cr
0 \qquad \hbox{otherwise}\;.\cr}
\label{core_prescription}
\end{equation}
where $M_{d, 1}$ is the mass of the dust in the first gas zone, $R_{2}$ is the
outer radius of the first gas zone, and $(u_{2}-u_{a2})$ is the velocity with
which grains arrive into the first zone from the second gas zone.  For
simplicity, we assume that the core has the same material density as the
grains, $\rho_a$. Therefore, the outer radius of the core is $R_{\rm core} =
(3M_{\rm core}/4\pi\rho_a)^{1/3}$. This radius becomes the new inner boundary
for the gas: as the core builds up, the inner radius of the first Lagrangian
mass zone becomes $R_{1} = R_{\rm core}$. As the radius of the heavy elements
core is initially orders of magnitude smaller than $R_2$ this does not lead to
any significant perturbations for the gas.

\subsection{Radiation from the forming core}\label{sec:core_radiation}

Accretion of grains onto the surface of the solid core generates accretion
luminosity equal to
\begin{equation}
L_{\rm acc} = \frac{d E_{\rm bind, c}}{dt} = \frac{G M_{\rm core}\dot M_{\rm
      core}}{R_{\rm core}}\;.
\label{Lacc}
\end{equation}
The radiative flux boundary condition is modified from the zero flux one used
in paper I, to read
\begin{equation}
F_{\rm rad}(R_1) = \frac{L_{\rm core}}{4\pi R_1^2}\;.
\label{fr_bound}
\end{equation}

\subsection{Fiducial parameters}\label{sec:fiducial}

The outcome of our grain sedimentation calculations depend on many parameters,
such as: the initial grain size $a_0$; grain mass fraction relative to gas,
$f_g$; turbulent mixing coefficient $\alpha_d$; maximum velocity $v_{\rm max}$
for grain growth; the opacity power law index $\alpha$; the opacity
coefficient $\kappa_0$; the mass of the first core, $\mfc$; and the initial (same as ambient) 
gas temperature $T_{\rm init}$. The parameter space is too large to
explore completely in one paper.

Fortunately, we find that some of the parameters listed above change the end
results very little except near regions delineating critical behaviour
changes. For example, the initial grain size is irrelevant in most cases as
small grains grow very quickly in our prescription due to Brownian motions. 
Grains also evaporate very quickly once gas
temperature exceeds $T_{\rm vap} \sim 1400$ K, and so their size is not
crucial in this case as well. Only when the other parameters of the problems
are such that the grain growth time, $t_{\rm gr}$, is about the vaporisation
time, $t_{\rm vap}$ (see paper I), the exact initial value of $a_0$ is
important.

We also find that there is a degeneracy in the role of some of the
parameters. As the interior temperatures of the first cores are typically of
order of 1000 K, the typical opacity of the first core is $\kappa \sim
\kappa_0 10^{2\alpha}$. A larger value of $\alpha$ can be compensated for by a
smaller value of $\kappa_0$ to yield a similar opacity through the first core,
although there may still be important differences between these cases near the
high-Z core or on the outskirts of the first core.

We shall therefore fix some of the less influential parameters at
``reasonable'' values, motivated by observations whenever possible, and also
try to avoid model degeneracies. In line with this, we set $f_g = 0.02$,
$T_{\rm init} = 10$ K, $\alpha = 1$, $a_0 = 1$ cm, $v_{\rm max} = 1$ m
s$^{-1}$. A logarithmic gas mass grid is used, so that the mass of grid cell
increases with index $i$ as $\Delta M_{i+1} = 1.08 \Delta M_{i}$, where $i =
1,2, ..., i_{max}$. We used $i_{max} = 170$ cells for most of the runs below,
which results in the mass of the innermost gas cell of $\sim 10^{-3} \mearth$
for $\mfc = 10 M_J$.

This leaves three parameters that we found to be instrumental in determining
the outcome of the calculations, and we shall vary these in a relatively broad
range: the opacity coefficient $\kappa_0$, the mass of the first core, $\mfc$,
and the turbulent mixing coefficient, $\alpha_d$.

\subsection{Simulations Table}\label{sec:table}

The parameters of the simulations selected for presentation in this paper, and
some of the more important results of these simulations are listed in Table
1. The first column is the simulation ``ID'', which starts with a letter M, H
or L. These stand to indicate the mass of the first core: ``H'' for heavy,
$\mfc = 20 M_J$; ``M'' for medium, $\mfc = 10 M_J$; and ``L'' for light, $\mfc
= 5 M_J$. The digit next to the first letter refers to the opacity coefficient
$\kappa_0$ of the simulation, such that $J_k$ satisfies $J_k = -
\log_{10} \kappa_0$. For example, $\kappa_0 = 0.01$ will have $J_k=2$ in its
ID. Similarly, the symbol $\alpha$ followed by a digit $J_\alpha$ signifies
the magnitude of the turbulent viscosity coefficient, $\alpha_d$, so that
$J_\alpha = - \log_{10} \alpha_d$.

The results part of Table 1 starts with two columns for two time variables,
$t_{\rm end}$ and $t_{\rm acc}$, both in units of $10^3$ years. The former is
the duration of the simulation. The simulations were run for as long as there
were an appreciable change in the results or until the simulations stalled due
to a vanishingly small time step. The latter time entry, $t_{\rm
  acc}$, is the time of the ``runaway solid core accretion phase''. The ``-''
symbol in this column indicates that the simulation never entered this phase
because grains vaporised earlier than they sedimented. The ``NA'' (not
applicable) entry refers to the cases when the core accretion is very gradual,
so that no runaway phase can be defined. This happens when turbulent mixing is
significant.

The last three columns in the Table relate to the masses of the first cores in
units of Earth masses. $M_{\rm core}$ is the final mass of the core, while
$M_{c1}$ and $M_{c2}$ are its masses at two intermediate times, $t=2\times
10^3$ and $t=10^4$ years, respectively. These two masses are given to indicate
the speed with which the cores are assembled.

\section{10 ${\bf M_J}$ first cores}\label{sec:10mj}

Here we present several calculations for first cores with mass of $10
M_J$. IDs of these runs start with letter ``M'' in Table 1. We shall first
look at several tests with the same level of turbulent mixing parameter,
$\alpha = 10^{-3}$, labelled $\alpha$3 in Table 1.

\subsection{Medium-high opacity case (M1{\bf $\alpha$}3)}\label{sec:largeK}

We start with a relatively high opacity case, $\kappa_0 = 0.1$. At the typical
first core's interior's temperature, we have $\kappa(T=10^3 \hbox{K}) = 10$
for this particular choice of $\kappa_0$. This is quite high but may well be
reasonable at least in the innermost parts of the first core when dust
sediments there.

\begin{figure}
\centerline{\psfig{file=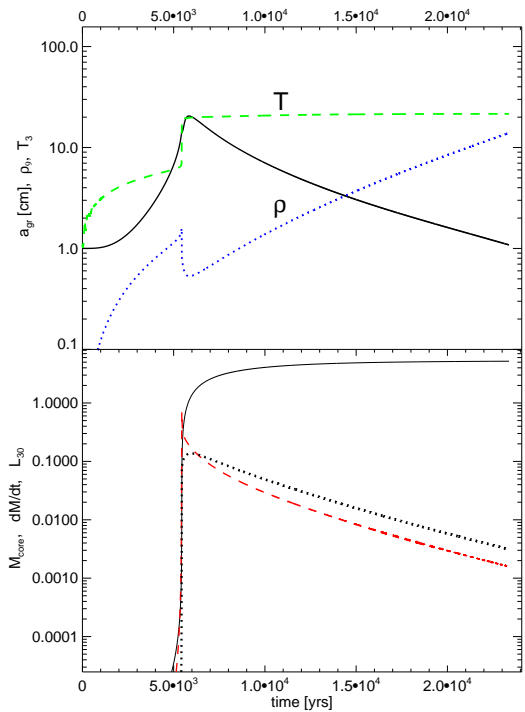,width=0.47\textwidth,angle=0}}
\caption{Results of simulation M1$\alpha$3 (see \S \ref{sec:largeK} and Table
  1). Upper panel: Grain size $a$ in cm (black solid curve), dust-averaged
  temperature $T_m$ (dashed green) in units of $10^3$ K, central gas density
  in units of $10^{-9}$ g cm$^{-3}$ (blue dotted). Lower panel: mass of the
  high-Z elements core, in Earth's masses (black solid), core's accretion rate
  in $10^{-2} \mearth$ yr$^{-1}$ (red dashed), and core's accretion luminosity
  in $10^{30}$ erg s$^{-1}$ (black dotted).}
\label{fig:hist_largeK}
\end{figure}

Figure \ref{fig:hist_largeK} shows the evolution of several key quantities as
a function of time for this calculation. The solid and dotted curves in the
upper panel of the figure show the grain size, $a$, and central gas density,
$\rho$, in units of $10^{-9}$ g cm$^{-3}$, respectively. The dashed curve is
the central gas temperature in units of 100 K. The solid curve in the lower
panel shows the mass of the high-Z core in units of Earth's mass, whereas the
dotted and the dashed curves show the core's luminosity (in $10^{30}$ erg
s$^{-1}$) and accretion rate, in $10^{-2} \mearth$ yr$^{-1}$.

The first $\sim 5 \times 10^{3}$ yrs of the simulation are familiar from paper
I. The grains must grow to $\sim 10$ cm size before a significant dust
sedimentation takes place. The evolution after that is marked by two stages:
the ``runaway'' stage during which the high-Z core grows very rapidly, almost
vertically in Figure \ref{fig:hist_largeK}, and the saturated growth stage
when the accretion rate, core's luminosity and grain size all drop with time.
This second stage is a testament to importance of feedback effects during the
core's growth. This can be deduced by noting the abrupt central temperature
increase at the end of the ``runaway phase'', when both the high-Z core mass
and luminosity increase sharply, and the corresponding central gas density
decrease. The temperature increase disrupts the hydrostatic balance
in the central regions, causing an outward gas expansion.

\begin{figure*}
\centerline{\psfig{file=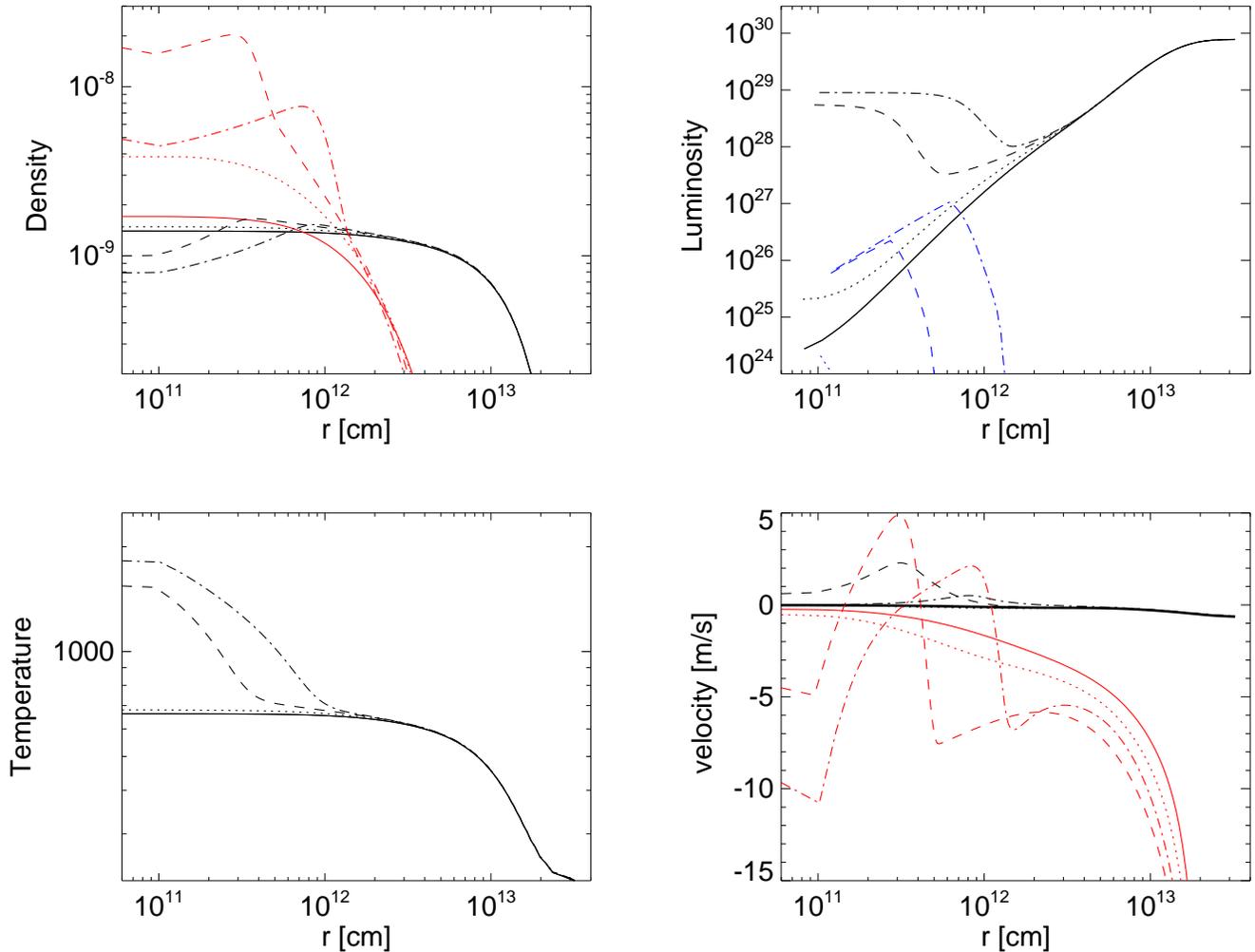,width=0.8\textwidth,angle=90}}
\caption{Radial structure of the first core during and just after the
  ``runaway growth phase'' in the simulation M1$\alpha$3 (see
  Fig. \ref{fig:hist_largeK}). Densities and velocities of gas and grains are
  shown with the black and red curves, respectively. The different style lines
  are for times $t=5386, 5417, 5446$ and 5469 years, for the solid, dotted,
  dashed and dash-dotted curves, respectively. The blue curves in the right
  upper panel show the convective luminosity which becomes relatively large
  after the high-Z core creation.}
\label{fig:sed_largeK}
\end{figure*}

These effects are seen in Figure \ref{fig:sed_largeK} that presents the radial
structure of the first core and the grain density distribution during and just
after the end of the runaway core growth phase. The panels show the gas and
the grain
densities, the gas temperature, the radiative (black curves) and the convective
luminosities (blue curves), and the gas (black) and the grain (red) radial
velocities. The convective luminosity is defined by $L_{\rm con} = 4 \pi R^2
F_{\rm con}$.

The first snapshot in figure \ref{fig:sed_largeK}, shown with the solid
curves, corresponds to the time when the grain density in the inner region is
equal to that of the gas, which corresponds to when we stopped  the simulations in paper
I. The mass of the high-Z core is only $\sim 3 \times 10^{-4} \mearth$ at
that point. As the grain density overtakes the gas density, the gas drag and
turbulent mixing resistance to grain sedimentation rapidly weaken. This is
seen in the inner part of the radial velocity curves sequence. As the inner
part of the ``grain sphere'' contracts due to its own self-gravity, the
accretion rate onto the high-Z core increases by orders of magnitude. The
luminosity of the solid core shoots up to about a tenth of that produced by
the whole first core (see the dashed and the dot-dashed curves in the upper
right panel of the figure).

The heat produced by the build up of the high-Z core strongly affects the
inner region of the cloud, heating up the gas there. The increased gas
pressure initiates a rarefaction wave that moves gas outward, as can be seen
in both the density and the velocity curves. It is likely that the grain
sedimentation would continue at this stage if it were not for the gas entropy
profile becoming strongly unstable to convection. The convective flux builds
up to its maximum $F_{\rm c,max}$ value allowed (blue curves in the upper
right panel) in the inner part of the cloud. Strong convective mixing drives
some grains in the inner region outwards (see the red velocity curves for grains). 
Note a very good correlation in the positions of the peak of gain velocity curves 
in the inner part of the embryo with the maximum in the convection luminosity. 
This confirms that the outward motion of grains is driven by convection in this region.

The second, also quite important way in which the high-Z core growth
self-regulates is vaporisation of grains. As the inner region of the
computational domain heats up, grain growth and evaporation start to compete
with each other. In fact the grain size decreases with time (upper panel in
fig. \ref{fig:hist_largeK}), although not particularly rapidly in this
simulation.

The existence of the two ways in which the high-Z cores can affect their
surroundings and hence limit the rate of their own growth suggests that this
feedback loop is quite robust qualitatively, despite quantitatively depending
on details of the gas-grain interaction prescriptions and the parameter
values.

These feedback effects preclude incorporation of all of the available
condensible material in the high-Z core. Indeed, as much as $\sim 20 \mearth$
of such material is present in the first core, but only a quarter of it gets
locked into the high-Z core. Clearly this result must be sensitive to the
opacity of the first core, as lower opacity would allow more heat to escape
the gas cloud, perhaps reducing the feedback effects.

\subsection{Medium opacity case}\label{sec:med_K}

\begin{figure}
\centerline{\psfig{file=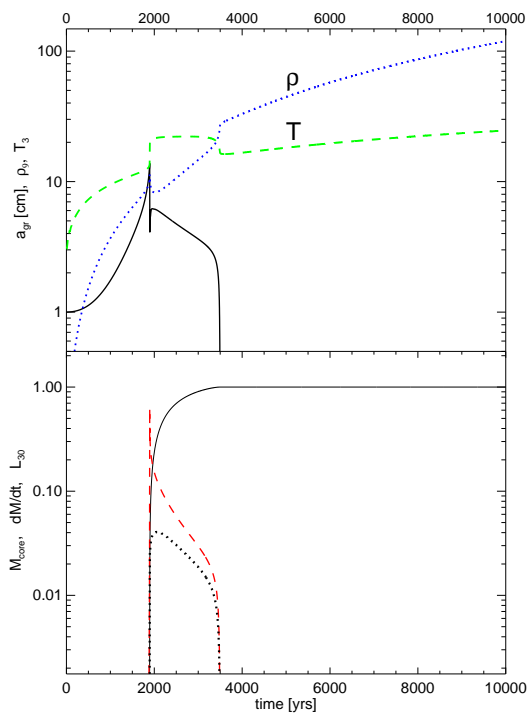,width=0.47\textwidth}}
\caption{Same as figure \ref{fig:hist_largeK} but for simulation
  M2$\alpha$3. Note the anti-correlation between the size of grains and
  the central temperature after the high-Z core is created. This is due to a
  feedback loop: grains power core's accretion luminosity that heats up the
  central region of the first core, but the grains can then be melted by if
  the gas becomes too hot.}
\label{fig:hist_medK}
\end{figure}

We now present calculation M2$\alpha$3 which is identical to M1$\alpha$3
except the opacity is reduced to $\kappa_0 = 0.01$. The resulting evolution of
the first core and the grain distribution is shown in Figure
\ref{fig:hist_medK}. Confirming the earlier made claim that embryo opacity is a 
key parameter of the problem, the
resulting change in the dust sedimentation outcome is significant. As the cooling time of the
first core decreases as $\kappa_0$ decreases, the cloud contracts more
rapidly. Higher gas density promotes faster grain growth. Therefore, grains
increase in size and sediment sooner in M2$\alpha$3 than they did in
simulation M1$\alpha$3. The runaway accretion phase begins earlier, at around
$2\times 10^3$ years. As the core accretion rate is higher, the feedback
effects of the growing high-Z core are more pronounced, and hence the rapid
growth of the core saturates at a lower core mass, and a lower luminosity
(cf. the lower panels of figures \ref{fig:hist_largeK} and
\ref{fig:hist_medK}). As in simulation M1$\alpha$3, the gas temperature near
the core is above vaporisation temperature for the grains, causing grain
melting and hence a plunge in the value of $a$ at the peak of the runaway
growth phase. In comparison to M1$\alpha$3, reduced opacity enables the gas
near the core to transfer the heat outward more rapidly. This allows a
further density and temperature increase near the core, evaporating the grains
further. A complete grain evaporation occurs at about 3,500 years in this
calculation. Grain accretion onto the high-Z core then ceases, allowing the
gas near the core to cool slightly. The contraction and the respective heating
of the gas near the core continues. This stops any further core re-growth.

The final mass of the high-Z core in this simulation is $M_{\rm core} = 0.9996
\mearth$. The fact that it so close to the Earth's mass is of course a complete
coincidence. Degrading the numerical resolution from $i_{max} = 170$ to
$i_{max} = 140$, and repeating the simulation M2$\alpha$3 leads to qualitatively
very similar results but the final core mass is $M_{\rm core} = 1.24
\mearth$. This implies that higher numerical resolution calculations would
yield somewhat lower masses for the final $M_{\rm core}$, but probably not by
more than a factor of 2. This is acceptable to us given large uncertainties in
parameter values. The decrease in the final core mass with an increasing
resolution is probably due to feedback effects becoming even more pronounced
as regions closer and closer to the core are resolved. We shall investigate
these issues further in the future.

\begin{figure*}
\centerline{\psfig{file=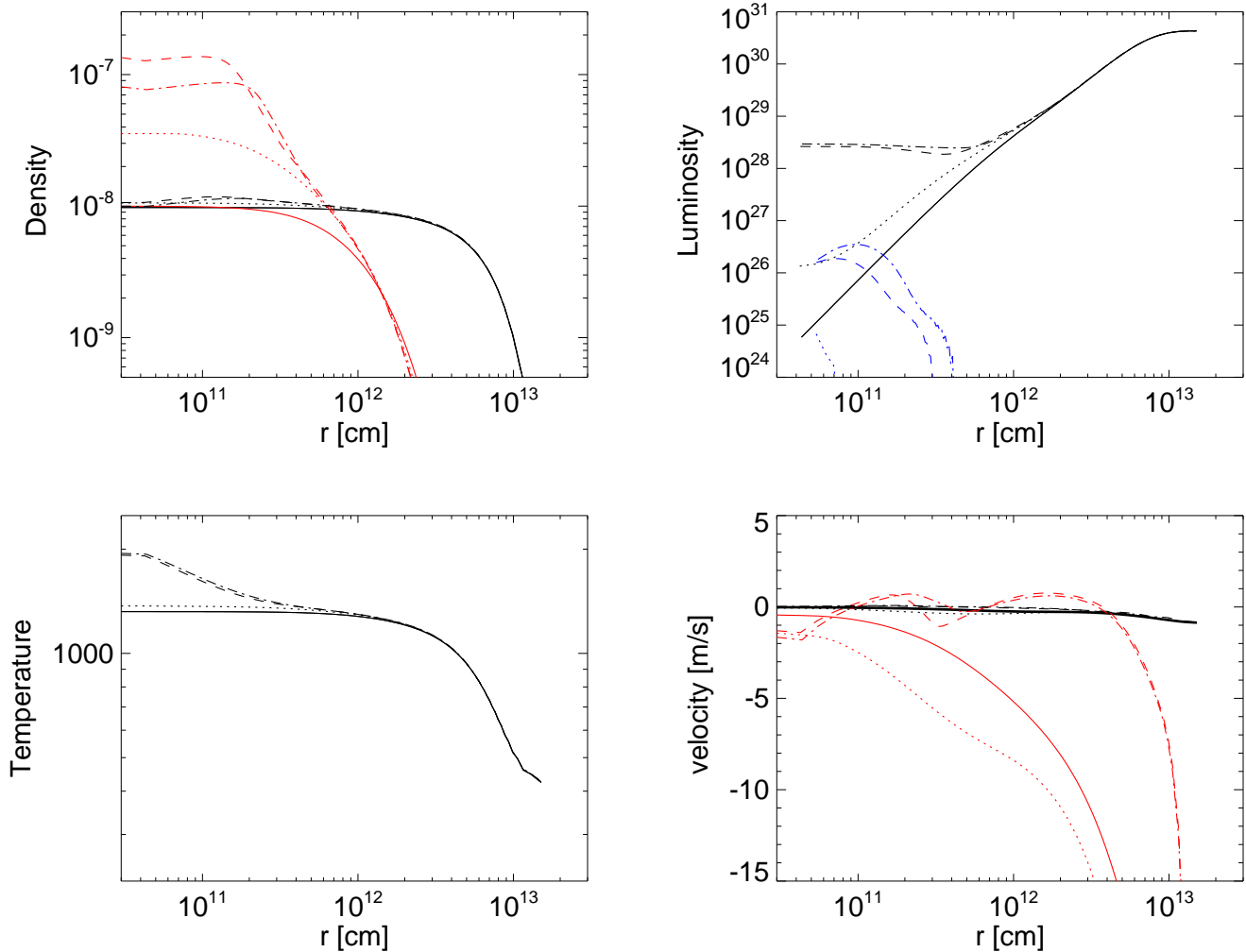,width=0.8\textwidth,angle=90}}
\caption{Same as figure \ref{fig:sed_largeK} but for simulation M2$\alpha$3
  (see also figure \ref{fig:hist_medK}). The snapshot times are $t=1880$, 1892,
  1902 and 1912 years.}
\label{fig:sed_medK}
\end{figure*}

Figure \ref{fig:sed_medK} shows the radial structure of the first cores at and
immediately after the runaway accretion phase for the simulation
M2$\alpha$3. There are many features similar to those seen earlier in figure
\ref{fig:sed_largeK}. Note a slightly lower core luminosity, which is in line
with the lower final core mass. It is also interesting that the general grain
sedimentation trend is reversed at ``large'' R, $R\sim 2 \times 10^{12}$ cm,
as grains there start to move outwards (see the lower right panel in figure
\ref{fig:sed_medK}). This is caused by a rapid decrease (by a factor of two)
in the grain size at the runaway accretion phase. The reduced size of grains
implies a smaller sedimentation velocity, and sedimentation is apparently
overpowered by turbulent mixing at these radii. The effect however does not
last for long as grains re-grow and sedimentation continues.

\subsection{Low opacity case}\label{sec:low_K}

Simulation M3$\alpha$3 probes the case of a yet smaller opacity,
$\kappa=0.001$. The cooling time of the first core in this case is very
short. Therefore, the results of this test are ``uninteresting'' -- grains
grow to $a\approx 1.5$ cm only to be evaporated at about $t \approx 200$
years. We present no figures for this calculation. Shortly after the grains
evaporate, the gas temperature approaches 2000 K, and the first core should
undergo the second collapse \citep{Masunaga00} and probably form a giant
planet mass object without a solid core.

\subsection{Very high opacity}\label{sec:very_highK}

\begin{figure}
\centerline{\psfig{file=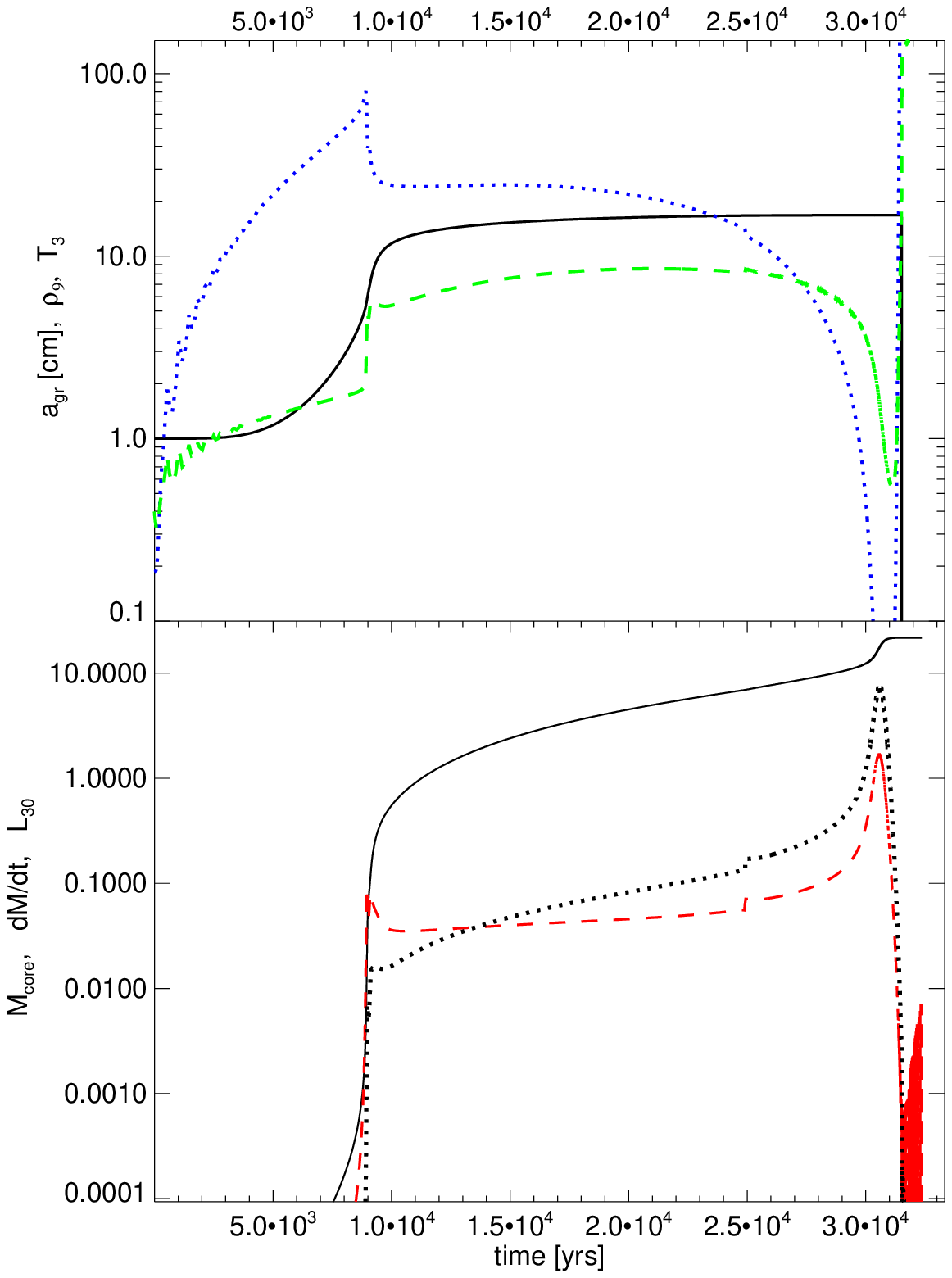,width=0.47\textwidth}}
\caption{Same as figure \ref{fig:hist_largeK} but for simulation
  M0$\alpha$3. In constraint to that figure, however, the density is plotted in
  units of $10^{-12}$ g cm$^{-3}$. Note the trough and then a spike in the
  temperature and density values near the end of the simulation. This marks
  building of a dense and hot atmosphere around the solid core (cf. figure
  \ref{fig:sed_hugeKa}).}
\label{fig:histk1}
\end{figure}

Simulation M0$\alpha$3 is the highest opacity case we consider here, with
$\kappa_0 = 1$. As the cooling time is very long for this simulation, the
first core does not contract very much before the grains sediment into the
high-Z core. The central density of the first core is much smaller than in the
previously investigated cases, and for this reason it is plotted in units of
$10^{-12}$ g cm$^{-3}$ in Figure \ref{fig:histk1}. As can be seen from the
figure, the simulation M0$\alpha$3 is very different from the lower opacity
cases studied so far.  The central temperature of the first core in
M0$\alpha$3 never becomes high enough to evaporate the grains, except at the
very end of the simulation. This is why the grain size (solid curve in the
upper panel of figure \ref{fig:histk1}) and the accretion rate onto the solid
core (red dashed curve in the lower panel of the figure) keep increasing through
and after the runaway accretion phase, whereas in M1$\alpha$3 and M2$\alpha$3
these quantities peaked during the runaway phase and subsequently dropped with
time.

Due to lower gas temperatures, the melting type of feedback does not take place
in M0$\alpha$3 before $t \sim 3\times 10^4$ yrs. Grains continue to sediment
and build up the central solid core at an increasing rate. The heat released
by the core is trapped near the core due to a higher opacity of the embryo. 
The only way for the gas in the inner part of the first core to absorb
the energy released by the high-Z core is to put it into a kinetic form. This
drives a strong outward expansion.

\begin{figure*}
\centerline{\psfig{file=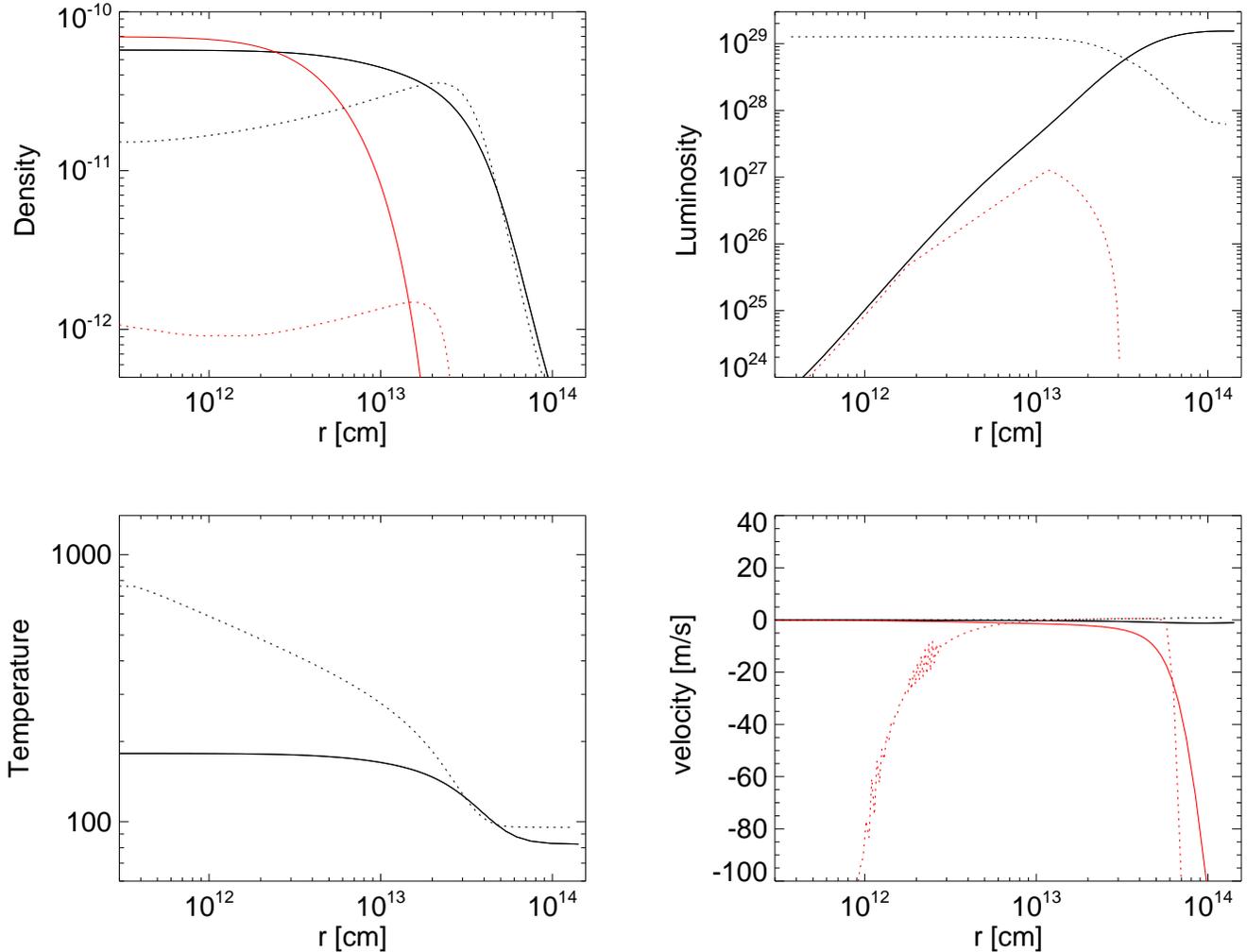,width=0.8\textwidth,angle=90}}
\caption{The radial structure of the first core near the end of simulation
  M0$\alpha$3 (see Fig. \ref{fig:histk1}). The solid and the dotted curves
  correspond to times $t = 8.5 \times 10^3$ and $t = 2.44 \times 10^4$ yrs,
  respectively. The red dotted curve in the top right panel shows the
  convective flux, saturated at the sonic value in the innermost part of the
  first core.}
\label{fig:sed_hugeKa}
\end{figure*}

Figure \ref{fig:sed_hugeKa} shows the radial structure of the first core near
the beginning of the high-Z core runaway accretion phase (solid curves) at $t=
8.5\times 10^3$ yrs, and at a later time, $t = 2.44 \times 10^4$ yrs (dotted
curves). Note an immense change in the luminosity curves (upper right panel of
the figure). While initially heat production was all due to the compressional
heating of the gas, at later times the energy production is dominated by the
grain accretion onto the high-Z core. The innermost region of the gas becomes
unstable to convection, and the convection flux then increases to the maximum
sonic value that we impose (\S \ref{sec:convection}). At the same time,
the convection flux is still very small compared with the radiative flux.

The increased luminosity of the high-Z core also forces the inner regions of
the gas to become much hotter, although not as hot as to evaporate the grains
(see the lower left panel of figure \ref{fig:sed_hugeKa}). The rarefaction in
the inner $\sim 10^{12}$ cm of the first core is significant, e.g., gas
density decreases by a factor of a few. Note that the motion of the gas and
the increased convective grain mixing does cause some outward radial motion of
the grains, although the innermost grain-dominated region becomes depleted
mainly due to grain accretion onto the high-Z core (see the red curves in the
upper left panel of the figure).

\begin{figure*}
\centerline{\psfig{file=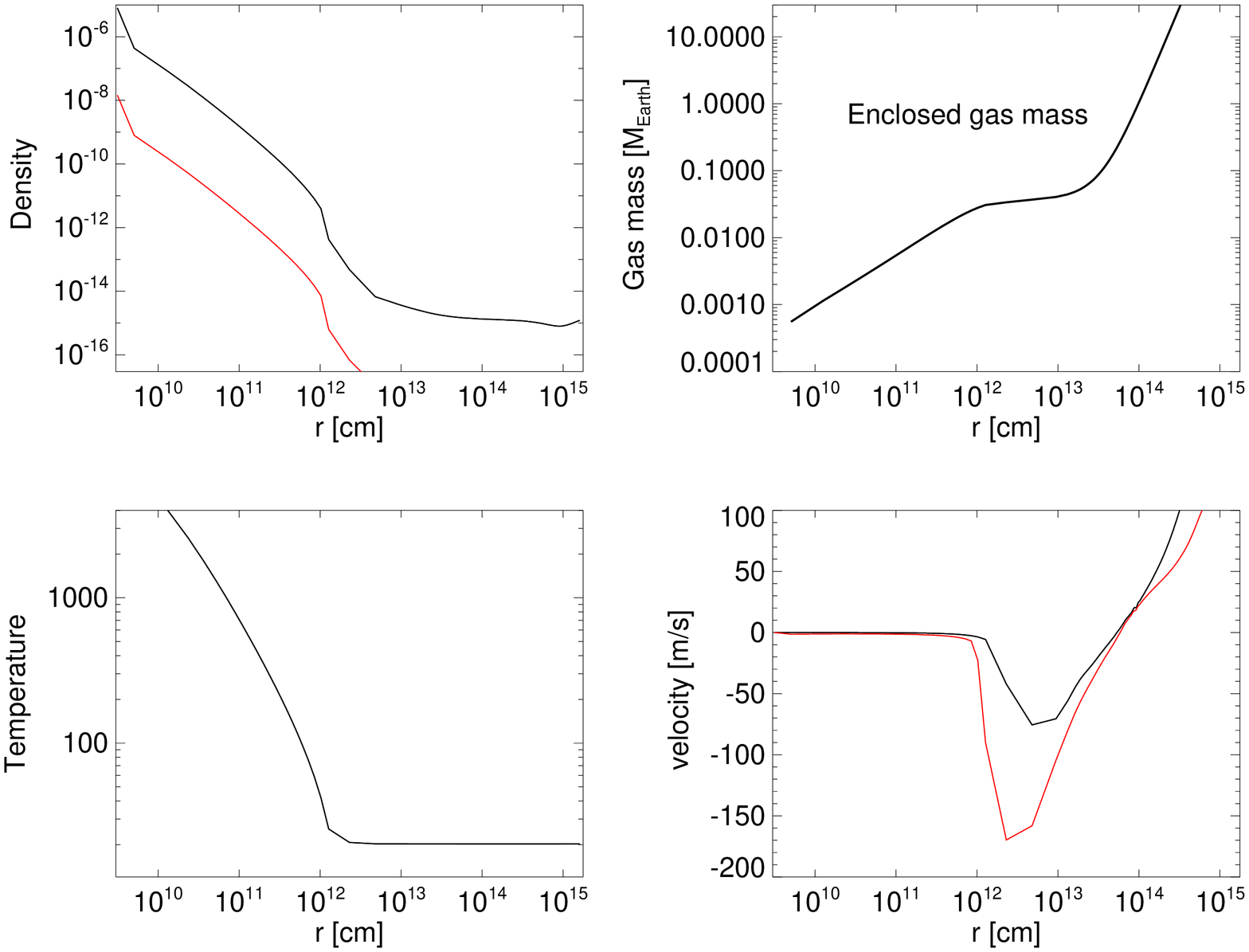,width=0.8\textwidth,angle=90}}
\caption{The radial structure of the first core near the end of simulation
  M0$\alpha$3, at time $t = 3.16 \times 10^4$ yrs. The upper right panel shows the
  enclosed gas mass in Earth's masses.}
\label{fig:sed_hugeK}
\end{figure*}

Further evolution of the first core is marked by a continuing energy release
from the solid core, which continues to inflate the hot bubble within the
first core until it actually breaks out -- unbinds most of the gas in the
outer parts of the first core.  This can be seen in Figure
\ref{fig:sed_hugeK}, which presents the radial structure at time $t=3.16\times
10^4$ yrs. Note the very high temperature and density at small radii and, on
the contrary, very low densities and a fixed temperature of $20$ K at large
radii. The latter is the minimum temperature imposed at the outer boundary for
this simulation (and for most others; normally this plays no role at all as
the temperature of the outermost zone is usually higher than 20 K). The radial
structure of the gaseous part of the first core in figure \ref{fig:sed_hugeK}
can be divided into three parts : (i) the dense and hot envelope tightly bound
to the high-Z core; (ii) the intermediate region at $10^{12} \simlt R \simlt
10^{13.5}$ cm; and (iii) the outer isothermal expanding shell. The last part
of the fist core is the most massive one; it is completely unbound and is lost
to the environment.

On the other hand, the region near the high-Z core can be called its
atmosphere as it is apparently bound to the core. It is likely that the
removal of the outer massive shell of gas actually promoted formation of the atmosphere
by letting the heat escape the inner region more easily. The atmosphere is not
massive at all: the upper left panel of figure \ref{fig:sed_hugeK} shows that
it is around $\sim 0.03 \mearth$, compared with the mass of the core, $M_{\rm
  core} \sim 20 \mearth$. Nevertheless, it is important as it indicates that
the solid cores formed inside the first cores can actually build up gaseous
atmospheres if right conditions exist.

\subsection{Nonlinear effects of turbulent mixing}\label{sec:imp_turb}

\begin{figure*}
\begin{tabular}{c|c}
{\psfig{file=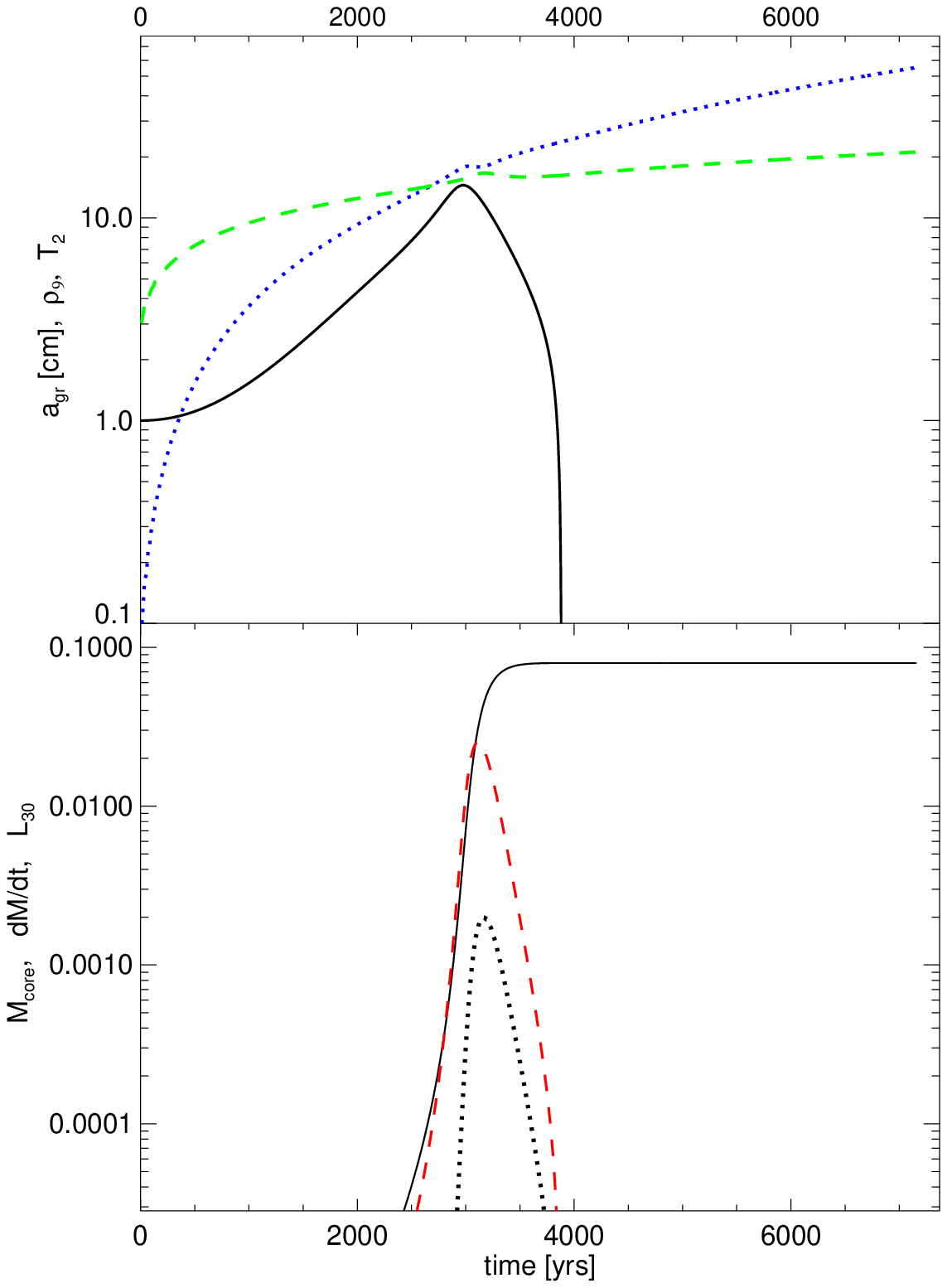,width=0.5\textwidth}}
&
{\psfig{file=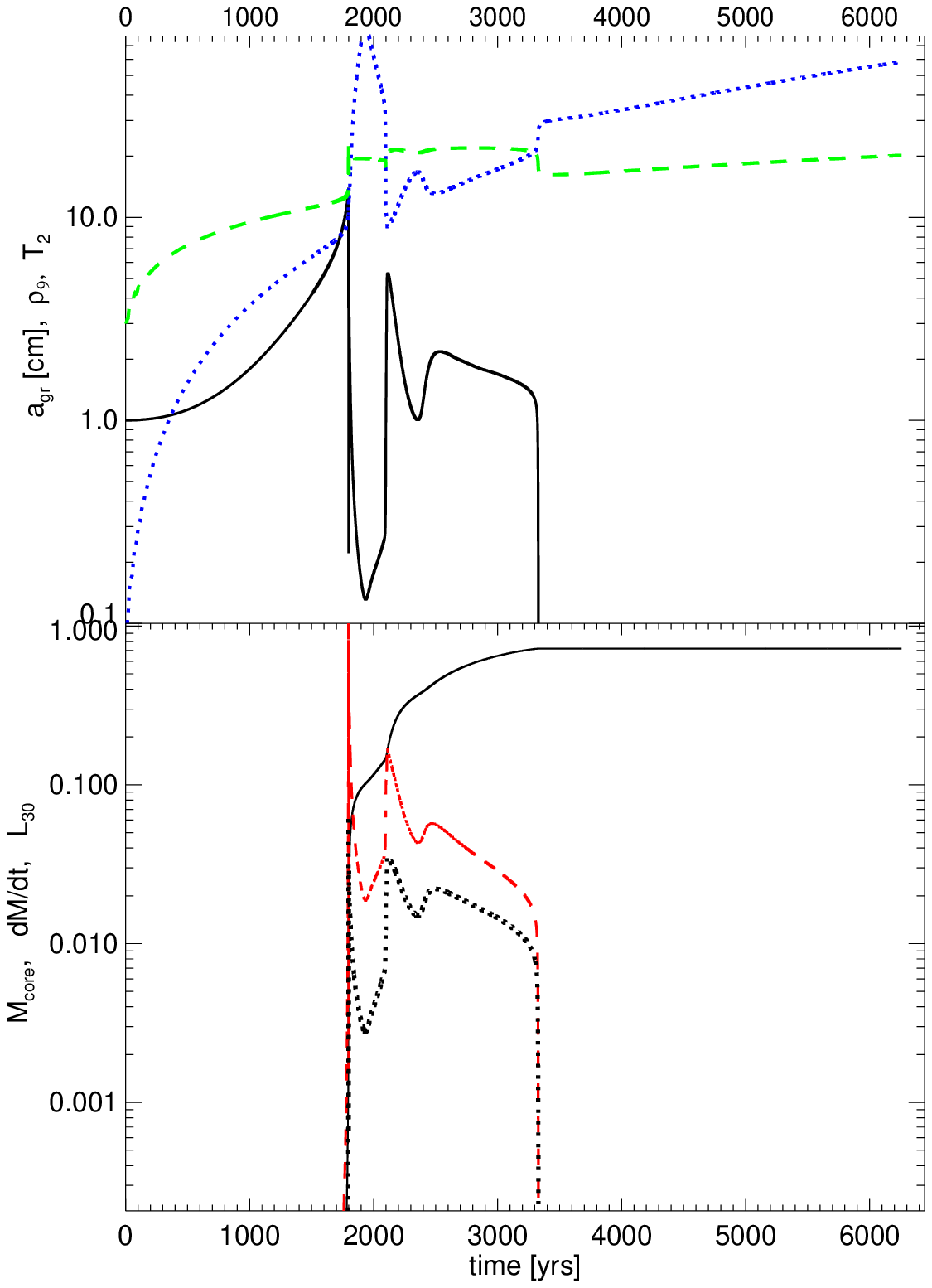,width=0.5\textwidth}}
\end{tabular}
\caption{Same as figure \ref{fig:hist_largeK} but for simulation M2$\alpha$2
  (Left panel) and M2$\alpha$4 (Right panel). Note that in both simulations
  the end mass of the solid core is lower than that in A2$\alpha$3, but for
  different reasons. See text (\S \ref{sec:imp_turb}) for a further
  explanation.}
\label{fig:hist_medKd2}
\end{figure*}

We found that the models can be surprisingly and non-monotonically sensitive
to the turbulent mixing parameter $\alpha_d$. Simulations M2$\alpha$2 and
M2$\alpha$4 are identical to M2$\alpha$3 except that $\alpha_d = 10^{-2}$ and
$10^{-4}$, respectively. The evolution of the first core properties of
simulation M2$\alpha$2 and M2$\alpha$4 is plotted in Figure
\ref{fig:hist_medKd2}, the left and the right panels, respectively.

Concentrating first on M2$\alpha$2, we observe that its outcome is drastically
different from that seen in figure \ref{fig:hist_medK}.  The larger turbulent
mixing delays grain sedimentation by only about 50\% (see $t_{\rm acc}$ in
Table 1), but the gas in the first core is hotter by this time. Importantly,
the gas temperature is now closer to the vaporisation temperature. As is clear
from the upper panel of figure \ref{fig:hist_medKd2}a, accretion luminosity of
high-Z core is only $\sim 10^{27}$ erg s$^{-1}$, butthis is sufficient to
push the gas near the core over the vaporisation temperature threshold. The
grains evaporate rapidly soon after this, and a further growth of the high-Z
core turns out impossible. The final mass of the core is $M_{\rm core} = 0.08
\mearth$. Note also that the accretion rate of grains on to the high-Z core is
smaller in M2$\alpha$2.

The simulation M2$\alpha$4 has a much lower level of turbulent mixing.
Recalling the trend from M2$\alpha$2 to M2$\alpha$3, e.g., the increasing mass
of the solid core with the decreasing turbulence parameter, one could expect
the solid core to be more massive in M2$\alpha$4 than that in M2$\alpha3$ ($1
\mearth$). However, the result is actually $M_{\rm core} \sim 0.7 \msun$. The
right panel of Figure \ref{fig:hist_medKd2} shows that the time evolution of
the central part of the cloud in this simulation is quite complex. The maximum
core accretion luminosity in this simulation actually exceeds that in
M2$\alpha$3 (compare the sharp red spikes in the luminosity curves for these two simulations). The
energy released by the core then melts the grains rapidly. While in
M2$\alpha$3 grain re-growth occured almost immediately after the accretion
luminosity drops, in M2$\alpha$4 the gas near the core is a little hotter and
hence the grains remain small for several hundred years. After the excess heat
finally leaves the central region, grain growth restarts. This apparently
leads to a second major core accretion episode at around 2200 years. The
intricate feedback loop steps in again, melting the grains somewhat, but this
time to the size of about 1 cm rather than $\sim 0.1$ cm. The grains are
eventually vaporised at time $t = 3300$ years, very similarly to the
simulation M2$\alpha3$. The core mass is then frozen at the final value of
$0.7 \mearth$.

Comparing the three simulations, it appears that the role of the turbulent
mixing is to smooth out the evolution of the grain size and the core accretion
rate. Indeed, we note that turbulent grain mixing, in prescription we are
using, may have both a negative and a positive effect on the solid core
growth. Prior to the formation of a relatively massive core, turbulent mixing
works only in one direction, that is opposing grain sedimentation. However,
once a sufficiently massive core forms, gravity in the inner regions become
dominated by the solid core itself rather than by the gas. The sedimentation
velocity of the grains then increases above the ``pre-core'' theoretical value
(paper I). The inner few zones of the core are then quickly emptied of the
grain material by accretion onto the solid core. This has the effect of
creating a positive rather than the usual negative gradient in the ratio of
$\rho_d/\rho$, e.g., a relative deficit of grains in the centre of the first
core. Turbulent diffusion mixing then reverses its effects and actually
increases the grain accretion rate {\rm near the core} while of course
continuing to impede grain sedimentation at larger radii.

\section{Dependence of results on the mass of the first cores}\label{sec:3d}

Having studied grain sedimentation and high-Z core building for $\mfc = 10
M_J$ in some detail, we shall now review the results of the less and more
massive runs shown in Table 1.

\subsection{``Heavy'' first cores}\label{sec:heavy}

The most massive first core case studied in the paper is $M_{\rm fc} = 20
M_J$, shown in Table 1 with IDs starting with letter ``H''. Only one value of
turbulent parameter is considered for these simulations, $\alpha_d =
0.001$. It is immediately obvious that these most massive cores are the least
promising sites of grain sedimentation, which is also in accord with the results of paper
I. Only the high opacity cases, $\kappa_0 = 0.1$ and 1, resulted in significant
high-Z cores. The physical reason for this is their already hot initial
state. These cores rapidly evaporate their grains, if allowed to cool at a
reasonably rapid rate.

The inefficiency of the process of dust sedimentation for these most massive
cores stands out clearly. The most massive solid core assembled, $M_{\rm
  core} = 12.2 \mearth$, represents only about a quarter of the condensible
high-Z mass available inside the gas clump for $M_{\rm fc} = 20
M_J$.

\subsection{``Low mass'' first cores}\label{sec:light_cores}

Finally we come to the lightest first core case we studied in this paper,
$M_{\rm fc} = 5 M_J$. Concentrating on the first four entries starting with
letter ``L'' in IDs in Table 1, one notices that these runs all produce high-Z
cores with masses above $1 \mearth$. The efficiency with which the grains are
locked into the central core reaches 100\% in L0$\alpha3$ and L1$\alpha3$.

Looking at further entries in the Table, we find the familiar trends with the
turbulent mixing parameter $\alpha_d$: higher values of it usually result in
smaller values for $M_{\rm core}$. The turbulence needs to be rather extreme
to prevent the formation of high-Z cores completely: a very high turbulence
run, $\alpha_d = 0.1$ (L0$\alpha1$), resulted in a core of $0.075 \mearth$
mass, albeit after a very long ($\sim 10^5$ years) time.

The physical reason for the resilience of dust sedimentation process 
in  lower mass cores is that the initial temperature in such is
low, e.g, $\simlt 100$ K (see the lower panel of Figure 1 in paper I). Both
cooling and vaporisation times in the low mass first cores are long, and hence
dust has a plenty of time to grow and sediment, even in the presence of
turbulence.

\section{Long term evolution}\label{sec:analytics}

\subsection{Expectations: Giant planets with solid
  cores?}

The solid core -- gas envelope structure we study here is essentially the same
as that studied by many authors in the context of the core accretion model for
giant planet formation \citep[e.g.,][]{PollackEtal96}. The difference is
mainly in the environmental factors by which we mean boundary conditions
``far'' from the solid core.  Furthermore, the solid cores that we study here
also grow by accretion of solids. We should thus expect that at least some of
the classical core accretion model results may pertain to the problem we
study.

One such result concerns the build up of a massive gaseous atmosphere around
the solid core. For low solid core masses, the atmosphere is in a hydrostatic
equilibrium, in which the gravitational force due to the solid core and the
gas own weight is balanced by the pressure gradient. As shown by
\cite{Mizuno80}, when the solid core mass increases and exceeds a critical
value, $M_{\rm crit}$, a hydrodynamical collapse driven by self-gravity of the
gas takes place. The gaseous envelope contracts to much higher densities. This
marks transition from a hydrostatic balanced atmosphere to a gas accretion
phase. The latter goes on as long as there is gaseous fuel supplied from
outside. Adding accretion of solids from the protostellar disc one arrives to
a very crude sketch of the core accretion model
\citep[e.g.,][]{PollackEtal96}.

There is a scope for a conceptual confusion here. The giant planet embryo is
itself a massive gaseous self-gravitating ``atmosphere'' around the solid
core. But the bulk of this gas is not bound to the solid core at all as the
mass of the latter is rather small. Initially, the gas pressure gradient is
nearly zero in the central part of the gas cloud since weight of the central
region is small (proportionally to $\sim R^3$). It is the weight of the layers
above that keep the innermost gas from expanding outward. Appearance of a
solid core in this constant pressure, constant density, environment, at first
has little effect as the gas pressure is too high. However, as the mass of the
solid core increases, there is a critical point as in the calculation of
\cite{Mizuno80} at which the gas around the solid core starts to collapse on
it and forms an atmosphere around it. This ``atmosphere within and
atmosphere'' is gravitationally bound to the solid core and itself. It is much
denser and becomes distinct from the giant embryo. In paper III we argue that
when the embryo is disrupted by tidal forces of the parent star, the inner
solid-core bound atmosphere may survive. The result is probably not unlike a
giant planet built in the core accretion model, although detailed calculations
beyond the scope of our paper are needed to characterise its chemical
composition and other properties. Nevertheless, one robust prediction is a
metal rich composition of a giant planet built inside a more massive giant
embryo gas clump: the abundance of dust in the central regions of the embryo
is enhanced by dust sedimentation.

\subsection{Numerical issues}

The most straightforward way to explore these ideas would be to study them
numerically. In the simulations listed in Table 1, there is only one example
of a bound gas atmosphere, the simulation M0$\alpha$3, and even then the
atmosphere weighs only $0.03\mearth$, which is about $0.001$ of the mass of
the high-Z core for that simulation.


We suspect that at least the more massive of the solid cores built up
in our simulations should be surrounded by massive gaseous atmospheres at
later times. Computational limitations of our method precluded us from
studying this interesting question in detail.

As explained in \S \ref{sec:general}, we need a hydrodynamics rather than
hydrostatics based approach to follow the evolution of the system during the
phases of a rapid gravitational collapse or the bounces driven by too much
accretion onto the solid core. Unfortunately this hydrodynamical approach also
makes it difficult to evolve the system for long periods of time after the
formation of the solid core if the core starts to build up a massive gaseous
atmosphere.  As an atmosphere starts to build up around the solid core, time
steps become very short, and the code basically stalls.

Being unable to follow the evolution of the system into the appropriate regime
numerically, we resort to analytical arguments and analogies to the classical
core accretion model results.

\del{Fortunately for us here, we can rely on previous calculations of massive
  atmosphere collapse in the context of core accretion models of giant planet
  formation \citep[e.g.,][]{Mizuno80}. In the following we make two opposite
  assumptions about the thermodynamics of the gas near the solid core, hoping
  to limit the critical core mass from below and from above.}

\subsection{Minimum critical (cold) solid core mass}\label{sec:crit_cold}

The temperature profile of the gas is the main uncertainty in our calculation
here. The lower the gas temperature, the smaller is the gas pressure gradient
for a fixed density profile. Therefore, by choosing appropriately a minimum
possible temperature that the gas near the solid core may have we should be
able to obtain the minimum required solid core mass for atmosphere collapse.

A simple way of doing this is to consider a giant embryo unperturbed by the
energy release from the solid core, setting $L_{\rm acc} = 0$. This is
satisfied by those of our models that had grains evaporated after the core
assembly (e.g., see the right panel of figure \ref{fig:hist_medKd2}). Further,
somewhat unrealistically, we shall assume that the gas remains isothermal even
in the presence of the solid core's potential. The result is nevertheless
interesting as a lower limit to the mass of the solid core at which a massive
gaseous atmosphere can be built.

With these ideas in mind, our treatment here echos that presented in section
6.1 of \cite{Armitage10} \citep[note that the derivation was first made
  by][]{Sasaki89}.  The hydrostatic balance equation for isothermal gas
surrounding the solid core is
\begin{equation}
c_s^2\; \frac{d\rho}{dR} = - \frac{GM_{\rm core}}{R^2} - \frac{GM_e(R)}{R^2} \;,
\label{hstat}
\end{equation}
where $c_s$ is the isothermal sound speed, $M_e(R) = \int_{R_{\rm core}}^{R} 4
\pi r^2 \rho(r) dr $ is the enclosed gaseous envelope mass inside radius $R$,
and $R_{\rm core}$ is the radius of the solid core. If a hydrostatic balance
solution exists, then the density given by equation \ref{hstat} should
smoothly join the unperturbed first core density at radii $R$ where the
gravity of the solid core is negligible.

We first consider the case when the gaseous envelope is much less massive than
the core, and thus we neglect the last term in equation \ref{hstat}. The
solution is then trivial:
\begin{equation}
\rho(R) = \rho_{\rm fc} \exp\left[\frac{GM_{\rm core}}{R c_s^2}\right]\quad
\hbox{for }\; R > R_{\rm core}\;.
\label{rho_atm}
\end{equation}
Note that $\rho(R) \rightarrow \rho_{\rm fc}$ at $R \rightarrow \infty$, as
needed.

The factor in the exponential can be very large near the solid core. In that
case it also drops off very quickly with the distance from the first
core. Hence, the mass of the atmosphere around the solid core is approximately
\begin{equation}
M_{\rm atm} \sim \frac{4\pi}{3} \rho_{\rm fc} R_{\rm core}^3
\exp\left[\frac{GM_{\rm core}}{R_{\rm core}
    c_s^2}\right]\;.
\label{mass_atm}
\end{equation}

This treatment is valid as long as $M_{\rm atm}\ll M_{\rm core}$, since in the
opposite case the weight of the atmosphere cannot be neglected in equation
\ref{hstat}, as we have done above. Self-gravity of the gas around the solid
core initiates a ``collapse'' of the gaseous atmosphere. The critical mass of
the solid core when the collapse happens is found by setting $M_{\rm atm} =
M_{\rm core}$. We call the corresponding core's mass a ``critical cold'' mass:
\begin{equation}
\frac{GM_{\rm cold}}{R_{\rm core} c_s^2} \approx \ln\left(\frac{\rho_c}{\rho_{\rm
    fc}}\right) = 20 \Lambda\;,
\label{mass_terr1}
\end{equation}
where $\Lambda\equiv \ln (\rho_c/\rho_{\rm fc})/20 \approx 1$ for $\rho_{\rm
  fc} \sim 10^{-11}$ g $cm^{-3}$. As the first core density enters
logarithmically, the result is weakly dependent on the value of $\Lambda$.
Eliminating $R_{\rm core} = (3M_{\rm core}/4\pi\rho_c)^{1/3}$, we have
\begin{equation}
M_{\rm cold} = \left(\frac{3}{4\pi \rho_c}\right)^{1/2}\;\left(\frac{kT}{\mu
  G}\right)^{3/2}\;\ln^{3/2}\left(\frac{\rho_c}{\rho_{\rm fc}}\right)
\label{mass_terr2}
\end{equation}
This expression is similar to that for Jean's mass, but with the density of the
solid core rather than the gas density inserted, and the logarithmic factor
$\Lambda$. Numerically,
\begin{equation}
M_{\rm cold} = 2.7 \, \mearth \; \frac{T_3^{3/2}}{\rho_p^{1/2}}\left(\frac{2.45}{\mu
  }\right)^{3/2}\; \Lambda^{3/2}\;.
\label{mass_terr3}
\end{equation}

This estimate is robust in the sense that the temperature of the unperturbed
first cores does not vary all that much. As found in paper I, by the time
grains sediment to the centre the temperature of the gas is typically between
a few hundred K and 1400 K. The upper value for the temperature range is set
by vaporisation of grains, whereas the lower one is simply due to a rapid
cooling of first cores: as we found in paper I, the first cores cool initially
very rapidly since they are spatially extended and only mildly optically
thick.

A correction downward to the minimum critical core mass is needed in the case
when the density of the dust around the solid core is comparable or higher
than that of the gas. One should then adjust the mean molecular weight in
equation \ref{mass_terr3} to reflect the metal-rich composition of material in
the inner region of the first core.

\subsection{Maximum critical (radiative) core mass}\label{sec:heating_effects} 

The isothermal envelope assumption that we used above allowed us to explore
one extreme of the problem at hand, e.g., when the envelope temperature is
completely unaffected by the presence of the solid core. An opposite extreme
assumption is that the immediate envelope around the solid core is much hotter
and denser than the giant embryo in which it is embedded, so that the presence
of the latter is a minor detail. This may nearly be the case if the luminosity
of the solid core is very high, e.g., comparable with the total cooling
luminosity of the giant embryo. This set of assumptions constitutes the
``radiative zero'' solution to the gaseous atmosphere structure
\citep{Stevenson82}.

The radiative zero solution also makes a prediction for the critical solid
core mass (see below), which we shall call $M_{\rm rad}$. In contrast to the
``critical cold'' mass derived above, $M_{\rm rad}$ is likely to overestimate
the critical core mass. The radiative zero solution assumes zero pressure and
density boundary conditions at infinity. In contrast, our envelope is
surrounded by material of a relatively high density and pressure. At large
distances from the solid core, gas density and temperature should join the
values of these quantities in the first core, rather than go to zero.  The
additional pressure and weight of the outer cooler envelope of the giant
embryo is certain to lower the critical mass of the solid core required to
initiate the collapse.

The central quantity in the radiate zero solution is the luminosity of the
solid core, which is derived from accretion of planetesimals. We thus need to
estimate the typical  accretion rate onto the core. The sedimentation time
scale of grain material, $t_{\rm sed}$ has been estimated in paper I, and is
usually between $10^3$ to a few $\times 10^4$ years. Assuming that $M_{\rm
  Z,c} \sim 20 \mearth$, and that all this solid core-building material falls
onto the core within $10^4$ years, we find the solid core accretion rate,
$M_{\rm core}$, is
\begin{equation}
\dot M_{\rm core} \sim 2\times 10^{-3} \mearth \;\hbox{yr}^{-1}\;.
\label{dot_est}
\end{equation}

Ikoma et al. 2000 updated and expanded the results of \cite{Stevenson82} for a
broad range of opacities in the envelope (which were assumed temperature
independent in contrast to our power-law dependence). In particular, they
provided a power-law fitting formula to the critical core mass. Using that
formula for a fiducial opacity coefficient value of $k_{\rm gr} = 0.01$, we
have
\begin{equation}
M_{\rm rad} \sim 25 \, \mearth \left(\frac{\dot M_{\rm core}}{2 \times 10^{-3}
  \mearth\hbox{yr}^{-1}}\right)^{q'} \left(\frac{k_{\rm gr}}{0.01 \hbox{cm}^2
  \hbox{g}^{-1}}\right)^{s'}
\label{mdot_ikoma}
\end{equation}
where $q' \approx s' \approx 1/4$. 

\subsection{Summary on critical core mass and further complications}

We have argued above that there should be a critical solid core mass, $M_{\rm
  crit}$, above which the initially tenuous gas atmosphere around the core is
unstable to self-gravity and should collapse. The collapse should build a {\em
  massive} gas atmosphere atop the solid core. The composition of the
atmosphere should be metal rich as the interior of the giant embryo is metal
rich due to dust sedimentation. This object inside the giant embryo may have
properties similar to giant gas and giant icy planets \citep[as pointed out
  by][]{BossEtal02}.

As far as the value of $M_{\rm crit}$, we suggested that it is limited from
the bottom by $M_{\rm cold}$ and from the top by $M_{\rm rad}$. For typical
parameters the range here is from a few to few tens of Earth masses. 

However, for the most massive embryos, say $\mfc \simgt 20 0$ Jupiter masses,
it is possible to get the gas to collapse gravitationally and form a compact
proto-planet without a solid core at all (see below), especially if dust
opacity can be made very small. In this case the embryo may reach vaporisation
temperature before dust sediments.

Our numerical code assumes that hydrogen is molecular, which is a very
reasonable assumption for the first cores before they undergo the second
collapse \citep[][]{Masunaga98,Masunaga00}. This assumption may be violated
when a dense and hot gas atmosphere builds up around the core, or when the
whole first core contracts sufficiently to surpass $T\approx 2000$ K. As with
collapse of first cores,  hydrogen molecules disassociation and then hydrogen atom
ionisation energy losses at even higher temperature may help gravitational
collapse of the atmosphere. 

Therefore, {\em if the first core is truly isolated } and is allowed to
contract in peace (rather than being tidally or irradiationally disrupted
after a finite amount of time, paper III), then the whole first core can
eventually collapse onto the solid core. This ``second'' wave of collapse is
completely analogous to that of the first cores in star formation
\citep{Larson69}, and is not at all initiated by the solid core. Hence {\em
  truly isolated} giant planet embryos could presumably contract into giant
planets with arbitrary small solid cores, including none.

\section{On assembly process of terrestrial planet cores}\label{sec:making}

In \S \ref{sec:analytics} we proposed that giant planets with solid cores may
form inside the giant embryos when the solid cores grow more massive than the
critical mass, $M_{\rm cr}$. We further argued that this part of the planet
formation picture may be quite similar to that worked out in the classical
core accretion schemes, albeit for different accretion rates of solid material
and boundary conditions ``far'' from the solid core.

In contrast, the earliest stages of the solid core assembly could not be more
different. We believe it is far ``easier'' to form massive solid cores inside
the giant embryos at large distances from the star than it is in situ at
$R\sim $ few AU in the ''standard'' bottom-up build-up scenario
\citep[e.g.,][]{Safronov69,Wetherill90}. The numerical treatment of dust
sedimentation inside the first cores ($=$ giant planet embryos) that we
employed in this paper may have obscured this important point, which we shall
discuss below in detail.

\subsection{The standard bottom-up scenario: the key role of
  planetesimals}\label{sec:bottom_up} 

The core accretion model --
\cite[e.g.,][]{Safronov69,Mizuno80,Wetherill90,IdaLin08}, also reviewed in
chapters 4-6 in \cite{Armitage10} -- stipulates that planet formation starts
on smallest scales, with dust initially growing inside the protoplanetary disc
into larger grains. This very first part of the picture is physically similar
to the model of \cite{Boss98} and what we assumed here, with dust growing by
gentle hit-and-stick collisions \citep{BlumWurm08}.  There is then a poorly
understood step of building planetesimals, which are solid bodies of $\simgt $
km size. Once planetesimals are assembled, ``planetary embryos'' -- $10^3$ km
sized bodies are built by collisions of planetesimals, and finally Earth mass
terrestrial cores are accreted by giant impacts of such embryos
\citep{Safronov69}.

Due to well known problems with planetesimal assembly authors used to urge a
leap of faith, arguing that if planetesimals had not been assembled somehow,
we would not be here, and that ``Nature knew how to do it''
\citep[e.g.,][]{Wetherill90}. This is clearly unsatisfactory. Most solid
matter outside the Sun in the Solar System is inside the cores of the giant
planets and also in the terrestrial planets. Thus the total mass of
planetesimals must have been at least equal to the total mass of the solids in
the Minimum Mass Solar Nebula \citep[MMSN, ][]{Weidenschilling77}, and hence
most of the solids in the disc (= ``nebula'') must have been reprocessed
through planetesimals.

Recent conceptual breakthroughs suggest that solids may be
concentrated into larger structures by instabilities and turbulence in
the disc \citep[e.g.,][]{YoudinGoodman05,JohansenEtal07,CuzziEtal08},
although it is too early to say that the planetesimal step assembly is
understood. Studies of collisional evolution of asteroid-like bodies
suggest that the initial size of planetesimals must be larger than the
classical $\sim few$ km size objects, and be in the 100 to 1000 km
size range \citep[e.g.,][]{MorbidelliEtal09}, thus increasing the
requirements for the planetesimal buildup models.

In any event, we shall now review the well known reasons why
terrestrial planet cores cannot be built in the CA model directly, and
argue that these difficulties are avoided if massive solid cores are
made inside giant planet embryos.

\subsection{Kepler shear: limiting the mass of fragments}\label{sec:shear}

The most direct route to ``large'' solid objects in the CA model is the
gravitational instability of a dense dust layer
\citep[][]{Safronov69,GoldreichWard73}.  It works best if there is no
turbulence in the gas, as turbulence provides support against gravitational
collapse. We hence first assume that turbulence is absent. Inside a
protostellar disc, Kepler shear limits the maximum mass of the clump of solid
material that can collapse into one body to \citep[e.g., \S 4.6.2
  in][]{Armitage10}
\begin{equation}
m_{\rm fr} \sim 4 \pi^5 \Sigma_{\rm sol}^3 G^2 \Omega_K^{-4}\;,
\end{equation}
where $\Sigma_{\rm sol}$ is the column density of solid material in
the disc, and $\Omega$ is the Keplerian angular frequency at that
location. At 1 AU, $\Sigma_{\rm sol} \sim 10$ g cm$^{-2}$
\citep{Weidenschilling77} . Thus, at the location of the Earth this
yields $m_{\rm fr} \simlt 10^{-9}$ Earth masses.

This shows that even in the most optimistic scenario one must start with very
small bodies and continue by building up larger bodies in collisions
\citep[e.g.,][]{MorbidelliEtal09}. The origin of the problem here is in the
differential shear of the disc that limits the largest possible wavelength of
gravitational instability \cite[e.g.,][]{Toomre64} to only $\sim 10^8$ cm (see
\S \ref{sec:prevent} below).

  Now, the mass of the solids inside of a bound fragment in the
  gravitationally fragmenting {\em gas} disc is much larger, e.g.,
\begin{equation}
m_{\rm fr} \sim f_g M_* \left(\frac{H}{R}\right)^3\;,
\label{msol}
\end{equation}
which for $H/R\sim 0.1$ yields at least a few Earth masses. This is
because (a) the parent gravitationally collapsing body is a gas disc,
not a dust one,
and (b), the Keplerian shear at 100 AU is $10^3$ times weaker than it
is at 1 AU.

\subsection{Kepler shear and turbulence: preventing
  collapse}\label{sec:prevent} 

Kepler shear has another undesirable effect on the potential gravitational
collapse of solid material inside the protoplanetary disc. To become
gravitationally unstable, the density of the layer must exceed the tidal
density, which is of the order of $10^{-7}$ g cm$^{-3}$ at 1 AU. Thus the
thickness of the layer must be less than $H_{\rm max} = \Sigma_{\rm
  sol}/\rho_{\rm t} \sim 10^8$ cm. In equilibrium, the thickness of the layer
of solid particles depends on their random, e.g., turbulent, velocities. These
must be less than $v_t \simlt (H_{\rm max}/R) v_K \sim 10^{-5} v_k = 30$ cm
s$^{-1}$. This is extremely small compared with the expected sound speed of
$\sim $ km s$^{-1}$. Such small turbulent velocities appear very unlikely
given the instabilities in the turbulent Ekman boundary layer \citep{Weiden80}
excited by the differential rotation of solids and the gas. Thus building up
solid bodies as small as $10^{-9} \mearth$ by gravitational collapse in the CA
model is not straightforward.

Now contrast this picture to the gravitational collapse of the dust sphere
settling inside the first core, a process analogous to the dust layer
gravitational collapse. As shown in paper I, gas pressure effects
become negligible when the grain sphere contracts to the ``grain cluster''
density $\rho_{\rm gc} = \rho_{\rm fc} f_g^{-1/2}$. Self-gravity of the grain
cluster at this point exceeds gas pressure gradients. The radial size of the
grain cluster is $R_{\rm gc} = \rfc f_g^{1/2}$, which is typically about $0.1
AU$, and the mass is $f_g \mfc$. The turbulent velocity required to keep the
grain sphere from collapse is
\begin{equation}
v_{\rm t,min} \sim \left(\frac{G f_g \mfc}{R_{\rm gc}}\right)^{1/2} \approx
f_g^{1/4} c_s\;,
\end{equation}
where $c_s = \sqrt{kT/\mu}\approx \sqrt{G\mfc/\rfc}$ is the sound speed of the
first core. As temperature of the first core is of order $10^3$ K, we find
\begin{equation}
v_{\rm t,min} \sim 0.6 \;\hbox{km s}^{-1}\; \left({f_g \over 0.01}\right)^{1/4}
T_3^{1/2}\;,
\label{vtmin}
\end{equation}
This velocity is roughly 30\% of the gas sound speed in the first core, and is
about 2000 times higher than the minimum turbulent velocity found for the dust
layer in the CA model. It seems very doubtful that the first core would
support such a strong turbulence. In contrast to the differentially rotating
Keplerian disc, the first core may be expected to be in a solid body rotation
with angular frequency characteristic of the disc at $R_p \sim 100$AU, e.g.,
3 orders of magnitude slower than at 1 AU.

These estimates are supported by our numerical results: even the highest
turbulent parameter run, $\alpha_d = 0.1$ (L0$\alpha$1), yielded a rather
massive solid core (cf. Table 1).

\subsection{Collisional buildup of planetesimals in CA model}\label{sec:coll_ca}

As the direct gravitational collapse to form (even such small bodies
as) planetesimals is complicated by turbulence in the CA model,
alternative ideas to form them via non-gravitational growth mechanisms
have been thoroughly explored \citep[e.g.,][]{DominikTielens97}; for a
recent review see \cite{BlumWurm08}. One could hope that the solids in
the disc continue to grow via collisional agglomeration as smaller
dust is believed to do.  However, there are two principal problems
with this hit-and-stick idea also. Firstly, due to drag from the gas,
the grains of different sizes are calculated to move at different
speeds; $\sim$ metre size objects should be colliding at velocities of
$\sim$ ten metres per second \citep{Weiden77}. Common sense and lab
experiments \citep{BlumWurm08} suggest that growth, i.e., sticking of
the collisions partners, is rather unlikely at these high speeds. The
second problem is frequently referred to as the {\em metre-size
  barrier}. In a non-turbulent protoplanetary disc, the $\sim $ metre
size objects must spiral radially inward, into the parent star, on
time scales of order a few hundred years, well before they could grow
to larger sizes \citep{Weiden77}.

\subsection{Planetesimals are not needed to build terrestrial planet cores inside
  giant embryos}\label{sec:no_planetesimals}

We find that neither the problem with too-energetic collisions for the
hit-and-stick growth nor the radial drift problem appear to exist if solid
bodies grew inside the protective environment of the giant embryo. The radial
drift problem is most obviously absent: solids migrate with the embryo rather
than individually.

Furthermore, we shall now show that solids that contribute the most to the
terrestrial planet cores inside the giant embryos are not planetesimals, and
not even boulders, but are meager pebbles. The collisions of such small bodies
are gentle. This resolved the issue with too fast collisions that tend to
fragment dust aggregates rather than bind collision partners together in the
CA model.


As we found in paper I, while grains are microscopic, turbulence keeps them
from sedimenting down very efficiently. The equilibrium reached between
sedimentation and turbulent mixing with the turbulent diffusion coefficient
$\alpha_d$ sets the radial scale of the dust distribution $H_d \sim \rfc
(\alpha_d \rho_{\rm fc}\rfc \lambda /\rho_a a^2)^{1/2}$ (see paper I). We
picked the ``large'' grain size limit here, when $a\gg \lambda \approx
4\;\hbox{cm}\; (10^{-9}/\rho_{\rm fc})$, the mean free path in the
gas. Requiring now that the equilibrium scale-height $H_d$ be smaller than the
grain cluster size $R_{\rm gc}$ -- at which point the grain cluster could
collapse due to self-gravity -- we infer that grains must be larger than
\begin{equation}
a_{\rm min} \sim 80\;\hbox{cm}\; \left(\alpha_{-3}f_{-2}^{-1} \rho_a^{-1}
\frac{\rfc}{1 \hbox{AU}}\right)^{1/2}\;,
\label{amin}
\end{equation}
where $\alpha_{-3} = \alpha_d/10^{-3}$. 

The estimate above predicts that the grains should grow to about 1 m size in
order for gravitational instability to overwhelm turbulent mixing. However, we
must keep in mind that the estimate is based on a number of approximations;
e.g., estimating the properties of the ``grain cluster'' based on the maximum
gas pressure gradient possible (paper I). Numerical experiments in the present
paper have shown that the ``runaway phase'' of core accretion, which occurs
when the grain cluster starts to collapse, takes place when the grains reach
dimensions of 10-20 cm (cf. figures in this paper), rather than 80 cm.

We can now estimate the sedimentation velocity of grains of this
characteristic size:
\begin{equation}
u_{\rm sed} = \frac{4 \pi G \Sigma_a R_{\rm gc}}{3 c_s} \;\frac{a}{\lambda} \;.
\label{vsed_max} 
\end{equation}
Since $\rfc/c_s\approx 1/\sqrt{G\rho_{\rm fc}}$, we can rewrite this as a
function of the gas density of the first core at the moment of the grain
cluster formation:
\begin{equation}
u_{\rm sed}  \approx 0.7 \;\hbox{m s}^{-1}\;
\left(\frac{a}{10\hbox{cm}}\right)^2 \rho_a f_{-2}^{1/2} \rho_{-9}^{1/2}
\label{vsed_max_num} 
\end{equation}
This estimate is consistent with our numerical results -- the grain
sedimentation velocity is below 10 m/s in the innermost part of the first core
identifiable as the ``grain cluster''.


One caveat here is that we have so far assumed a non-rotating giant
embryo. However, comparing the stopping time of a grain defined as
$u_a/|du_a/dt|$, one finds that only grains larger than $10^3$ cm or more
would be in the ``perturbed Kepler orbit'' regime \citep{Weiden77} in which
the particle moves essentially on its Keplerian orbit experiencing little drag
from the gas. The particle's angular momentum suspends it some distance away
from the cloud centre, and it spirals in slowly due to gas drag forces.
This implies that small particles dominating the grain sedimentation process
are in the opposite regime, e.g., approximately following the rotation of the
gas. As long as the embryo is not completely supported by rotation, the local
rotation speed is lower than the circular orbit value. Thus small grains
cannot be supported by rotation against gravitational collapse, and our
spherically symmetric treatment should be reasonably representative of slowly
rotating giant embryos.

Having said this, we note a sort of reincarnation of the ``metre-size
barrier'' problem for dust in rotating giant planet embryos -- even larger
than 1 metre dust grains are going to be lost into the centre of the embryo --
but that is not a problem at all. The grains sinking into the centre of the
cloud are not lost into the star but presumably join the solid core there,
therefore they are not ``lost'' in any sense.

\subsection{Summary on direct gravitational collapse}

Summarising these points, it appears that a direct build up of massive
terrestrial-like cores is possible inside the first cores (giant planet
embryos) due to a favourable (spherical) geometry and low shear. One
interesting point of the present picture is that the building blocks of the
solid cores of terrestrial planets are grains of $\sim 10$ cm size rather than
planetesimals of km or more size. While numerical 3D simulations, well beyond
the scope of our paper, are needed to model the amount of mass put into bodies
of different sizes, there is no clear theoretical need to put much mass into
the planetesimal population inside the embryo at any time. This is in contrast
to the core accretion model where terrestrial planets are made of
planetesimals.  We hence expect (pending future confirmation by numerical
simulations) that the relative amount of mass in asteroids and comet-like
bodies compared to that in terrestrial cores is much smaller in our model than
traditionally assumed.

Finally, there is no loss of solids problem as in the disc geometry -- any
solids lost into the centre of the giant embryo actually contribute to the
growth of the solid core there.

\section{Discussion}

\subsection{Summary of numerical results}

We have argued that clumps formed by the GI in proto-planetary discs at $R\sim
100$ AU should be similar to that of the first cores \citep{Larson69}. Using a
two-fluid radiative hydrodynamics code, we found that the dust inside the gas
clumps may grow and settle into a core if the gas remains cooler than the
grain vaporisation temperature. The high-Z elements core may become massive
($\sim 10 \mearth$) if feedback processes are not too severe. Our results are
more restrictive than analytical estimates of \cite{Boss98,BossEtal02}. On the
other hand, the maximum masses of the solid cores that we find are larger than
those of \cite{HelledEtal08,HS08}.

As already indicated in the Introduction, the main difference with the latter
authors appears different initial conditions for the gaseous clumps. For
example, all of the $M > 5 M_J$ mass clumps in Table 1 of \cite{HS08} are
unable to form any solid cores because they are already hotter than $T =1400$
K at birth. In contrast, the initial temperature of a $\mfc = 5 M_J$ clump in
our model is only about 60 K (cf. equation \ref{tvir_core}). These clumps do
cool initially very rapidly, increasing in temperature to hundreds of K, but
their increased density allows for an accelerated grain growth too (see
fig. 1-3 in paper I). We also found that our initial conditions are stable to
convection, and thus convective grain mixing does not significantly impede
grain sedimentation. Convection becomes important once a massive and
radiatively bright solid core forms, but by that time grains may be large and
the central region dominated (by density) by the grains, thus the effects of
convection remain subdued compared with the results of
\cite{HelledEtal08,HS08}.


\subsection{Astrophysical implications}\label{sec:astro}

We discuss observational implications of these results in a follow-up paper
(paper III) where we take into account additional processes not discussed
here: radial migration of the giant planet embryos and their Tidal and
Irradiative Evaporation (TIE) when the embryos come too close to the parent
star. Therefore here we only briefly mention the implications directly
following from the results presented here and in paper I.

We believe that grain sedimentation inside the giant planet embryos at tens to
a few hundred AU distances from the parent star is a viable process by which
to form terrestrial mass solid cores. We argued in \S \ref{sec:making} that a
direct gravitational collapse of $\sim 10$ cm-sized solids in the central
region of the gas clump can make massive solid cores directly without the need
to go through objects of intermediate sizes (such as planetesimals and
Moon-sized objects). Neither turbulence nor the high speed collisions (absent
in TIE in the stages before the grain cluster collapse) appear to be serious
obstacles to the gravitational collapse of the solid component.

We also argued that massive gaseous atmospheres around the solid cores should
start building up once the latter reach a critical mass which is in the range
of a few to a few tens Earth masses. The arguments here are similar to the
classical core accretion picture for building the giant planets
\citep{Mizuno80,PollackEtal96} except that the gas is accreted not from the
disc but from the surrounding gaseous envelope of the first core. This
potentially can make a metal rich (as the interior of the first core is such)
giant planet with a solid core. Removal of the outer metal poor gaseous
envelope, discussed in the follow-up paper, could then produce classical giant
planets such as Jupiter.

In addition to the tidal disruption and irradiative evaporation of the
envelope by the parent star, gaseous clumps may self-unbind themselves due to
an excessive heat release by the solid cores within them if envelope opacity
is high. These effects may be even more important for rotating first cores
\citep[rotation is expected for first cores formed in a disc,
  e.g.,][]{Levin07}. Indeed, the amount of energy needed to unbind the core is
then reduced. Furthermore, following the suggestion of \cite{BossEtal02}, the
gaseous envelope of the first cores may be removed by a strong ionising
radiation of a passing OB star. This is clearly more realistic for first cores
as these are initially very fluffy objects \citep[see also][]{WZ04}, but more
work is needed here. We plan to address the issue in our future work.

The processes discussed here may create terrestrial mass solid planets at
large distances from the parent star. In addition, the escape velocity form
first cores is less than a few km s$^{-1}$. Collisions of such rotating clumps
with each other may perhaps unbind a fraction of the solid material there,
contributing to the population of solids at these large radii \citep[see
  also][]{ClarkeLodato09}.

Finally, we note that the proposed route to formation of terrestrial and giant
planets is somewhat a of a hybrid between the two current competing models,
e.g., the gravitational instability and the core accretion models. In TIE the
planet formation process begins with gravitational instability (gas clump
formation, grain sedimentation and core growth) but continues with as in core
accretion (accretion of gas onto the solid core).

\subsection{Shortcomings of the calculations}\label{sec:open}

Despite using a relatively sophisticated two-fluid radiation hydrodynamics
approach, our treatment here is still quite far from being sufficient. In
particular, hydrogen molecules dissociation and hydrogen ionisation should be
taken into account in the future to allow higher resolution studies of the
innermost region near the solid core. Introducing several different species of
grains with different vaporisation temperature is desirable, as is allowing
the grain size to vary from bin to bin, instead of assuming one size as we
have done here. Turbulent and convective mixing for grains may be improved.
Furthermore, we have neglected effects of rotation of the first core here.
Finally, the dust opacity, independent and fixed in our calculation, should of
course evolve as dust grows and sediments, but it is hard to see how a
self-consistent and yet not overly complicated treatment of these effects
could be done at the present.

We expect these effects to modify the results of our dust sedimentation
calculations. Nevertheless, we feel that our exploratory calculations are
interesting in showing the possible range of behaviour in the given problem.

\subsection{Connection to previous ideas and work}\label{sec:previous}

As spelled out in the Introduction, our work is an extention and an update of
the \cite{Boss98,BossEtal02} ideas. We find that giant planet embryos born at
$R \sim 100$ AU should be similar in their initial properties to the first
cores. These cooler and fluffier gas configurations present a perfect
birthplace for massive solid cores. In paper III we argue that giant planet
embryos migrate inward and get tidally disrupted there. This hypothesis is
similar to the ideas of Boss and co-authors, although the exact protoplanetary
disc setting is different.

There is also some similarity between our ideas and that of
\cite{RiceEtal04,ClarkeLodato09}, who have recently argued that some fraction
of planetesimals may form inside gas pressure maxima provided by spiral arms
in self-gravitating protoplanetary discs at $R\simgt 30$ AU. These spiral arms
are stable for about a rotation, during which time the density enhancement in
the arms, compared to the mean density in the disc, may be significant if the
cooling time in the disc is short enough. The location of the marginally
stable {\em non-fragmenting} region of the disc is somewhat inside the
fragmenting region that we discuss \citep[see Fig. 1 in][]{ClarkeLodato09}.
Our study is thus an extension of these ideas on the larger disc scales, where
spiral arms do fragment, and where the density enhancement is much greater
than unity.

Furthermore, we note that, having submitted papers I and III, and having
almost finished the final draft of the present paper we discovered a set of
even more closely related ideas on formation of terrestrial planet cores
inside giant planet embryos in a recent paper by \cite{BoleyEtal10}. In
addition, \cite{BoleyDurisen10} present numerical simulations of fragmenting
gaseous disc which includes a population of dust grains and also argue for
grain sedimentation inside the giant planet embryos. While we have so far
concentrated on the internal evolution of individual giant embryos,
\cite{BoleyDurisen10} paid more attention to the interaction of the embryos
with their parent gas disc. These completely independent results can be cited
as a direct support of our ideas.

\section{Conclusions}\label{sec:conclusions}

We argued that gas clumps spawned by gravitational instability in massive
young proto-planetary discs may be excellent sites  for grain growth and
sedimentation, in agreement with the ideas of \cite{Boss98,BossEtal02}.  This
may allow a direct gravitational collapse of the grain component of the
central part of the giant planet embryo. This process spawns terrestrial mass
solid cores without going through km-sized bodies (\S \ref{sec:making}).  It
was also suggested that, similarly to the core accretion scenario of giant
planet formation, there is a critical solid core mass at which the gas
atmosphere around the solid core collapses to much higher densities. We
suggest that this process may create metal rich giant planets with solid cores
inside the massive gas clumps. In a companion Letter we show that these clumps
should migrate inwards where they are disrupted by tidal and irradiative
evaporation, releasing the inner planet. We believe this new hybrid scenario for
planet formation is promising and should be explored further.

\begin{table*}
\caption{Parameters of the simulations (see \S \ref{sec:table} for
  detail). ID: id of the simulations. The mass of the first core, $\mfc$, is
  in Jupiter's masses; $k_0$ is the opacity coefficient in the opacity law
  given by equation \ref{kappa0}; $\alpha$ is the power law index in that law;
  $f_g$ is the grain mass fraction of the core; $t_{\rm end}$, in $10^3$
  years, is the duration of the simulation; $t_{\rm acc}$ is the time at which
  the runaway accretion on the high-Z core sets in, in units of $10^3$ yrs;
  $M_{\rm core}$ is the core mass at $t=t_{\rm end}$; $M_{c1}$ and $M_{c2}$
  are same but at times $t = 2 \times 10^3$ years and $t = 10^4$ years,
  respectively.}
\begin{center}
\begin{tabular}{|l|c|c|c|c|c|c|c|c|c}\hline
ID & $\mfc$ & $k_0$ & $\alpha_d$ & $f_g$  & $t_{\rm end}$& $t_{\rm acc}$ (10$^3$
yrs) & $M_{\rm core}$ & $M_{c1}$ & $M_{c2}$ 
\\
                                  \hline
\hline
M0$\alpha$3 & 10  & 1 & $10^{-3}$ & 0.02  & 31.5 & $9.1$ & 21.6 & $\sim 0$ & 0.7\\ \hline
M1$\alpha$3 & 10  & 0.1 & $10^{-3}$ & 0.02  & 24 & $5.0$ & 5.2 &  $\sim 0$ & 4.0\\ \hline
M2$\alpha$3 & 10  & 0.01 & $10^{-3}$  & 0.02  & 10 & 1.9 & 1.0 & 0.2 & 1.0 \\ \hline
M3$\alpha$3 & 10  & 0.001 & $10^{-3}$ & 0.02  & 1.5 & -- & $10^{-6}$ & -- & -- \\ \hline
M2$\alpha$2 & 10  & 0.01 & $10^{-2}$ & 0.02  & 7.2 & 3.0 & 0.08 &  $10^{-6}$ & 0.08\\ \hline
M2$\alpha$4 & 10  & 0.01 & $10^{-4}$ & 0.02  & 6.4 & 1.8 & 0.7 & 0.12 & 0.7\\ \hline
\hline

H0$\alpha$3 & 20  & 1 & $10^{-3}$ & 0.02  & 30. & 12.5 & 12.2 & $3\times
10^{-5}$ & $0.0015$ \\ \hline
H1$\alpha$3 & 20  & 0.1 & $10^{-3}$ & 0.02  & 10.5 & 5.2 & 1.5 & $2.8\times
10^{-5}$ & $1.5$ \\ \hline
H2$\alpha$3 & 20  & 0.01 & $10^{-3}$ & 0.02  & 3. & -- & $10^{-6}$ & $10^{-6}$
& $10^{-6}$ \\ \hline
H3$\alpha$3 & 20  & 0.001 & $10^{-3}$ & 0.02  & 3. & -- & $\sim 0$ & $\sim 0$
& $\sim 0$ \\ \hline
\hline

L0$\alpha$3 & 5  & 1 & $10^{-3}$ & 0.02  & 39.7 & $\sim 9.0$ & 10.8 & $2\times
10^{-5}$ & $0.15$ \\ \hline
L1$\alpha$3 & 5  & 0.1 & $10^{-3}$ & 0.02  & 20. & 6.5 & 10.8 & $1.5\times
10^{-5}$ & 3.4 \\ \hline
L2$\alpha$3 & 5  & 0.01 & $10^{-3}$ & 0.02  & 32 & 2.4 & 6.2 & $6 \times 10^{-5}$ 
& 6.1 \\ \hline
L3$\alpha$3 & 5  & 0.001 & $10^{-3}$ & 0.02  & 2.2 & 0.7 & 1.2 & 1.2
& 1.2 \\ \hline
L0$\alpha$1 & 5  & 1 & $0.1$ & 0.02  & 135 & NA & 0.075 & $2\times
10^{-5}$ & $10^{-4}$ \\ \hline
L0$\alpha$2 & 5  & 1 & $0.01$ & 0.02  & 46 & NA & 0.005 & $2\times
10^{-5}$ & $3 \times 10^{-4}$ \\ \hline
L1$\alpha$4 & 5  & 0.1 & $10^{-4}$ & 0.02  & 22. & 5.2 & 10.8 & $5\times
10^{-6}$ & 4.4 \\ \hline

L2$\alpha$2 & 5  & 0.01 & $0.01$ & 0.02  & 14.4 & 4.2 & 3.1 & $2.2 \times 10^{-5}$ 
& 2.9 \\ \hline
\end{tabular}
\end{center}
\label{table1}
\end{table*}

\section{Acknowledgments}

The author acknowledges illuminating discussions with Seung-Hoon Cha, Richard
Alexander, Phil Armitage and thanks Giuseppe Lodato for pointing out the
\cite{ClarkeLodato09} paper, and discussions of massive proto-stellar discs.
Theoretical astrophysics research at the University of Leicester is supported
by a STFC Rolling grant.


\label{lastpage}

\end{document}